\documentclass[showpacs,reprint,aps,pra]{revtex4-1}
\usepackage{hyperref}
\hypersetup{
colorlinks=true,        
linkcolor=blue,          
citecolor=blue,         
urlcolor=blue            
}
\usepackage[utf8]{inputenx}

\usepackage{subcaption}

\usepackage{amsmath}
\usepackage{amsfonts}
\usepackage{amssymb}
\usepackage[normalem]{ulem}

\usepackage{graphicx}
\graphicspath{{./images/}}

\begin{document}

\title{Size-dependent commensurability and its possible role in determining the frictional behavior of adsorbed systems}

\author{Paolo Restuccia$^{1}$}
\author{Mauro Ferrario$^{1}$}
\author{Pier Luigi Silvestrelli$^{2}$}
\author{Giampaolo Mistura$^{2}$}
\author{Maria Clelia Righi$^{1,3}$}
\email{mcrighi@unimore.it}
\affiliation{$^{1}$ Dipartimento di Scienze Fisiche, Informatiche e Matematiche, 
Universit\`a di Modena e Reggio Emilia, Via Campi 213/A, I--41125 Modena, Italy\\ 
$^{2}$ Dipartimento di Fisica e Astronomia, Universit\`a di Padova, via Marzolo 8, I--35131, Padova, Italy\\
$^{3}$ CNR-Institute of Nanoscience, S3 Center,  Via Campi 213/A, I--41125 Modena, Italy}


\begin{abstract}

Recent nanofriction experiments of xenon on graphene revealed that the slip
onset can be induced by increasing the adsorbate coverage above a critical value,
which depends on temperature. Moreover, the xenon slippage on gold
is much higher than on graphene in spite of the same physical nature of the interactions.
To shed light on these intriguing results we have performed molecular dynamics
simulations relying on \textit{ab initio} derived potentials.
By monitoring the interfacial structure factor as a function of coverage and temperature,
we show that the key mechanism to interpret the observed frictional phenomena
is the size-dependence of the island commensurability.
The latter quantity is deeply affected also by the lattice misfit,
which explains the different frictional behavior of Xe on graphene and gold.

\end{abstract}

\maketitle

\section{introduction}

Submonolayer islands of rare gas atoms adsorbed on crystal surfaces offer an excellent platform to address friction
at crystalline interfaces. In the submonolayer range ($0 < \theta  < 1$, where $\theta$ is the coverage)
and at low temperatures, adsorbate phase diagrams are well known to display phase-separated two-dimensional (2D) solid islands, usually incommensurate with the surface lattice, coexisting with the 2D adatom vapour~\cite{Bruch2007}.
The inertial sliding friction of these islands can be accurately measured using a quartz crystal microbalance (QCM)~\cite{Krim88}. The condensation of a film on the QCM electrodes is signaled
by a decrease in the resonance frequency. Any dissipation taking place at the solid–film interface is instead detected
by a decrease in the corresponding resonance amplitude. From the shifts in the resonance frequency and amplitude
of the QCM one can calculate the slip time $\tau_s$~\cite{Bruschi2001}:
it represents the time constant of the exponential film velocity decrease due to a hypothetical sudden stop of the oscillating substrate. Very low $\tau_s$ means high interfacial viscosity and, in the case of a film rigidly locked
to the substrate, $\tau_s$ goes to zero.

Many different friction phenomena of sub-monolayer islands have been studied
using the QCM, including:
the dynamic depinning of Kr on gold~\cite{Bruschi2002},
the impact of substrate corrugation on the sliding friction levels of adsorbed films~\cite{Coffey2005},
the existence of an onset coverage for sliding of Ne islands on lead~\cite{Bruschi2006},
the dynamical sticking of solid helium on graphite~\cite{Hosomi2009},
the nanofriction of various adsorbates on superconducting lead~\cite{Dayo98,Renner2001,Pierno2010},
and the superlubricity of Xe islands on copper~\cite{Pierno2015}.

The nanotribological studies based on the QCM, which involve clean, well defined interfaces,
represent an ideal situation where simulations can combine with experiments 
to unravel fundamental aspects of friction. 
Recent findings include the scaling behavior of
the island edge contribution to static friction,\cite{Varini2015}
the effect of non linear dynamics,\cite{Manini2016} 
the influence of the adsorbate-substrate interaction on friction,\cite{Zhang2011} 
the size dependence of static friction,\cite{Reguzzoni2012}
the importance of the domain nucleation,\cite{Reguzzoni2010} 
the nature of thermally activated island creep\cite{Persson2003} 
and, more in general, the energy dissipation in monolayer films
sliding along substrates.\cite{Smith96,Vanossi2013,Krim2012}

The nanotribological studies based on the QCM, which involve clean, well defined interfaces,
represent an ideal situation where simulations can combine with experiments
to unravel fundamental aspects of friction.
Recent findings include the scaling behavior of
the island edge contribution to static friction~\cite{Varini2015},
the effect of non-linear dynamics~\cite{Manini2016},
the influence of the adsorbate–substrate interaction on friction~\cite{Zhang2011},
the size dependence of static friction~\cite{Reguzzoni2012},
the importance of the domain nucleation~\cite{Reguzzoni2010},
the nature of thermally activated island creep~\cite{Persson2003}
and, more in general, the energy dissipation in monolayer films
sliding along substrates~\cite{Smith96,Vanossi2013,Krim2012}.

Nanoparticle manipulation~\cite{Dietzel07} is also emerging as an invaluable tool
to investigate the frictional properties of nanoaggregates on surfaces,
both experimentally and by theoretical simulations.
Recent findings, complementary and closely related to the QCM ones,
include the size-scaling of structural lubricity~\cite{Wijn12,Dietzel13,Dietzel14},
the influence of contact aging~\cite{Sharp16},
and thickness promoted lubricity for metallic clusters~\cite{Guerra16}.

Recently, Pierno \textit{et al.} have measured the sliding time of Xe on graphene using a QCM
technique between 25 and 50 K~\cite{mistura} and found that the Xe monolayer is pinned to
the graphene surface at temperatures below 30 K. At T = 35 K, the Xe film starts to slide
for a coverage of 0.45 monolayer (ML). The critical coverage for sliding decreases with
increasing temperature afterwards. By comparison, under similar conditions, the slip time of Xe on the bare gold surface
is about twice higher than that on graphene, showing a lower depinning temperature of 25 K.
This is counterintuitive since it is believed that the hexagonal graphene layer is smoother for
adsorbate than gold surfaces. The understanding of the physical mechanisms behind this puzzling
observation is the subject of this theoretical investigation, in which we perform a comparative
study of xenon on graphene and gold substrates. By means of classical molecular dynamics,
relying on \textit{ab initio} derived potentials, we simulate the particle diffusion on the two
substrates and monitor the commensurability of the growing islands as a function of coverage and temperature.

\section{Systems and methods}
\label{sec:method}

\begin{figure}[htpb]
\begin{center}
\includegraphics[width=0.48\linewidth]{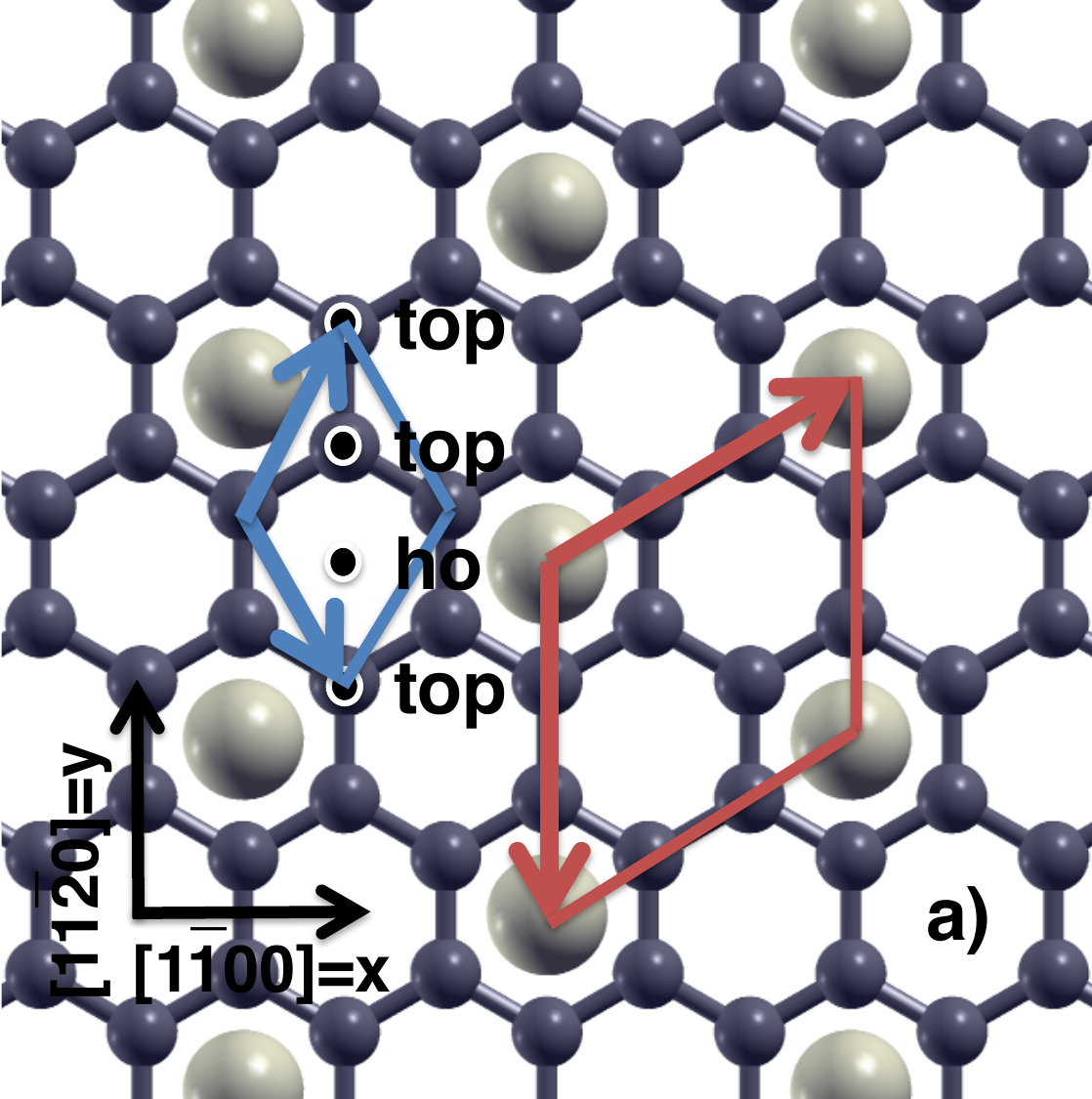}
\includegraphics[width=0.48\linewidth]{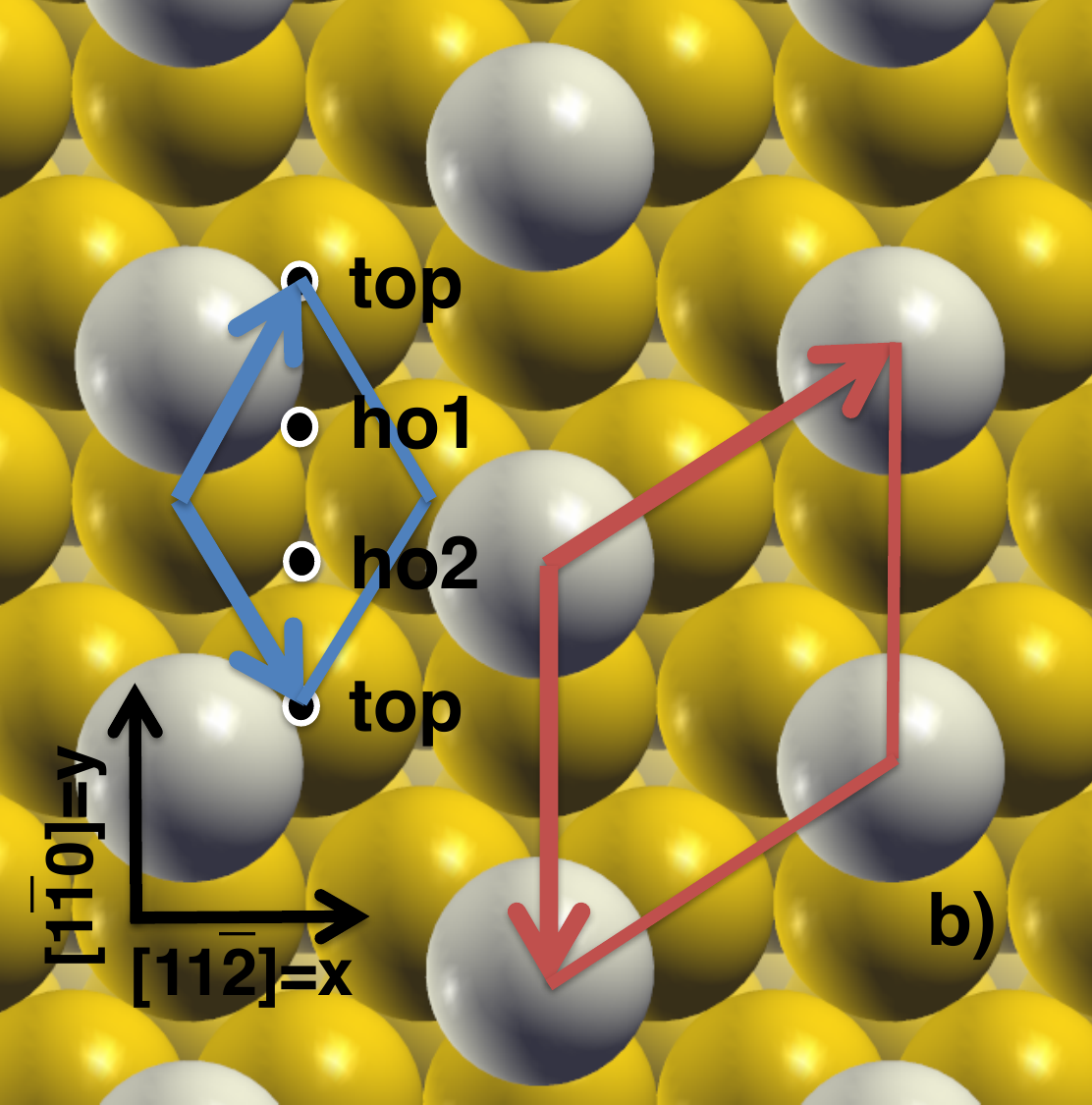}
\end{center}
\caption{Xenon atoms arranged in a $\left( \sqrt{3} \times \sqrt{3} \right) R 30^{\circ}$ lattice on a graphene layer (left panel) and Au(111) surface (right panel). The unit cells of the hexagonal lattices are indicated by arrows, while the labels indicate the location of high symmetry sites.}
\label{fig:xe}
\end{figure}

At full monolayer (ML) coverage and temperature below 70 K, the xenon adatoms form a $\left( \sqrt{3} \times \sqrt{3} \right) R 30^{\circ}$ sublattice on the graphene layer~\cite{pussi} (GL) (left panel of Fig. Fig.~\ref{fig:xe}). The mismatch
between the lattice parameters of an isolated Xe monolayer and the graphene $\left( \sqrt{3} \times \sqrt{3} \right) R 30^{\circ}$ sublattice is very small, equal to $-2.8$\%.
The same adsorption configuration for xenon-on-gold (Xe/Au)
(right panel of Fig.~\ref{fig:xe}) is characterized by a much larger mismatch of about $+13.9$\%. 

We sample the potential energy surface (PES), which describes the
interaction between the Xe adlayer and the substrate as a function of their relative position,
by means of \textit{ab initio} calculations based on density functional theory (DFT).
The results are used to optimize the parameters of the interaction potentials
subsequently employed in molecular dynamics simulation systems
with a large number of atoms.

We perform DFT calculations using the Quantum Espresso
package~\cite{pw}, initially within the local density approximation (LDA)
for the exchange correlation functional. Afterwards, to better describe the
van der Waals (vdW) interactions, the numerical approach has been improved
by using the rVV10 density functional~\cite{rvv10}.
rVV10 denotes a more efficient version of the original VV10
scheme~\cite{Vydrov} and is based on a nonlocal correlation functional
which provides an accurate description of van der Waals effects.
rVV10 is expected to perform better than, for instance, PBE-D~\cite{grimme}
which is instead a scheme where the semilocal PBE functional is modified
by simply adding an empirical potential. This is particularly true for metal
and semimetal systems, such as the substrates considered in our study,
where the electronic charge density is relatively delocalized, so that empirical,
atom-based vdW corrections turn out to be inadequate.

The ionic species are described by pseudopotentials
and the electronic wavefunctions are expanded in plane waves. For the Xe/GL (Xe/Au)
system a kinetic energy cutoff of 30 Ry (25 Ry) is used to truncate the expansion
on the basis of preliminary calculations optimizing the lattice parameter of graphene (Au).
We use periodic supercells with $( \sqrt{3} \times \sqrt{3} )$ in-plane size
and vertical edge $42$~\r{A}  ($46$~\r{A}) long. The \textbf{k}-point sampling of the Brillouin zone is realized
with a $8 \times 8 \times 1$ ($6 \times 6 \times 1$) Monkhorst-Pack grid.\cite{monkhorst}

The interaction energy per particle, $E$, for a Xe layer on the substrate is calculated as
$E = (1/N)[E_{tot} - (E_{sub} + NE_{Xe})]$,
where $E_{tot}$ is the total energy of the adsorbate system, $E_{sub}$ is the energy of
the isolated substrate and $ E_{Xe}$ is the energy per particle in the isolated xenon layer. All these systems
are described with the same $( \sqrt{3} \times \sqrt{3} )$ supercell.
The supercell used to model the Xe/GL system contains $N$ = 1 Xe atom,
while the supercell used to model the Xe/Au system contains $N$ = 2 Xe atoms
adsorbed on the two opposite surfaces of the slab at sites
with the same symmetry on the opposite sides. This reduces
spurious dipole–dipole interactions between periodically
repeated images. The Au slab is 7 layer thick.

Classical molecular dynamics simulations (MD) are performed by using the LAMMPS~\cite{plimpton}
computer code. The adopted force fields are constructed by putting together the following set of interactions.
The xenon–xenon interaction is described by the pair potential proposed by Tang and Toennies~\cite{tang}

\begin{equation}
   V_{T} (\mathbf{r}) = A_{T} e^{-b_{T} \mathbf{r}} - \sum_{n=3}^N f_{2n}(b_{T} \mathbf{r}) \frac{C_{2n}}{\mathbf{r}^{2n}} \text{,}
\end{equation}

\noindent where $\mathbf{r}$ is the distance between two xenon atoms, $f_{2n}$
is a damping function for the attractive part of the potential that can be expressed in terms
of the incomplete gamma function,
$f_{2n}(x) = 1 - \frac{\Gamma(2n+1,x)}{2n!}$.
The adopted numerical values for the $A_{T}$, $b_{T}$ and $C_{2n}$ coefficients are
the same as in the original paper~\cite{tang}.
This potential provides an improved description, with respect to the Lennard Jones potential,
of both the repulsive part of the interaction (with the use of the exponential
function instead of the too steep $\mathbf{r}^{-12}$ term) and the attractive part
(including the higher order dispersion $\mathbf{r}^{-8}$ and $\mathbf{r}^{-10}$
terms in addition to the standard $\mathbf{r}^{-6}$ one).

The xenon–carbon interaction is described by the
functional form of the Buckingham potential~\cite{Buckingham38}

\begin{equation}
   V_B (\mathbf{r}) = A_B e^{-b_B \mathbf{r}} - \frac{C_B}{\mathbf{r}^6} \text{;}
\label{eq:buck}
\end{equation}

\noindent where the $A_B$, $b_B$ and $C_B$ coefficients are optimized in order to reproduce 
the \textit{ab initio} PES, as detailed in Section~\ref{subsec:PES}. 
A distance cutoff of 12~\r{A} is used for both the Tang-Toennies 
and the Buckingham potentials. 

The interactions among the carbon atoms of graphene are modelled by
the second generation REBO potential~\cite{rebo,stuart}.
To avoid large out-of-plane
deformations of the graphene layer and mimic the experimental conditions,
we model the presence of a (rigid) substrate
underlying the graphene layer by means of the following analytic potential:

\begin{equation}
   V(x,y,z) = C_0(x,y) e^{-z C_1(x,y)} - \frac{C_2(x,y)}{z^4} \text{,}
\end{equation}

\noindent where $(x,y,z)$ are the coordinates of a carbon atom on the substrate, $C_i(x,y) = C_i^{max} - (C_i^{max} - C_i^{min}) u(x,y)$ for $i=0, 1, 2$ 
 and $u(x,y) = \frac{2}{9}[3 - 2 \cos \theta_x \cos \theta_y + \cos 2 \theta_y]$,
$ \theta_x = \frac{2\pi x}{a}$ and $ \theta_y = \frac{2\pi y}{a\sqrt3}$. 
The numerical values for the $a$ and $C_i^{min/max}$ coefficients are reported in Ref.~\citenum{reguzzoniPES}.

Finally, the interaction of xenon with the Au(111) surface is modeled by the three-dimensional 
periodic function~\cite{modelxe} 

\begin{equation}
	V(x,y,z) = A_0(x,y) e^{-z \cdot (A_1(x,y))} - \frac{A_2(x,y)}{z^3} \text{,}
\label{eq:potxe_onmetal}
\end{equation}

where $A_i(x,y) = A_i^{top} + \frac{9}{8} (A_i^{ho} - A_i^{top}) u(x,y)$ for $i=1, 2$,
with $u(x,y) = \left[ 3 - \sum_{\mathbf{g}} \cos{\mathbf{g} \cdot \mathbf{r}} \right]$
and the summation running over the first three $\mathbf{g}$ vector of
the reciprocal lattice and $A_0(x,y)$ is given by

\begin{equation}
	A_0 (x,y) = A_0^{top} \exp\left[ \frac{\left( A_1(x,y) - A_1^{top} \right)}{A_1^{ho} - A_1^{top}} \ln \left( \frac{A_0^{ho}}{A_0^{top}} \right) \right] \text{.}
\end{equation}

\noindent This function is able to accurately describe
the peculiar features of the PES for rare gases on metals~\cite{modelxe,modelxepress}. In particular, its
``anticorrugation'', which originates from the on-top site preference
for rare gas adsorption rather than the hollow site~\cite{dasilva,dasilvametal}.
This feature cannot be reproduced by pair-wise potentials, such as
the Lennard-Jones potential, which favor adsorption at the highest coordinated sites.
The numerical parameters entering the $A_i(x,y)$ periodic functions are
determined by fitting \textit{ab initio} adsorption energies calculated
for several configurations of the xenon atom on the metal surface,
as described in the next section.

MD calculations are performed using periodic boundary conditions. The orthorhombic cell used to model
the Xe/GL (Xe/Au) system has $15.25 \; \text{nm} \; \times 14.67 \; \text{nm} \; \times 2.00 \; \text{nm} $
($18.17 \; \text{nm} \; \times 17.48 \; \text{nm} \; \times 2.004 \; \text{nm} $) dimensions.
We consider three different Xe coverages on both the GL and Au(111) surfaces.
The adatom coverage is calculated assuming that y = 100\% corresponds to a complete
Xe ML with an areal density of 5.94 atoms per nm$^2$, i.e., considering a nearest neighbor
distance among Xe atoms of 0.44 nm. This choice is consistent with that adopted in the analysis of the QCM
experiments~\cite{mistura}. The simulated coverages y = 11\%, 22\%, and 44\%
correspond, thus, to 147, 294, and 588 adparticles in the Xe/GL
system, and to 208, 416, and 832 particles in the Xe/Au system,
1328 and 1886 being the total numbers of atoms necessary to
realize a 100\% coverage of the two-dimensional cells used in the two systems.

The MD simulations are started from initial conditions constructed by assigning to 
the xenon atoms random, non-overlapping, positions sampled from a uniform distribution on the surface plane
in the MD cell, and random velocities sampled from the Maxwell distribution at the temperature $T_o=25$~K. 
The system is then thermalized at $T_{\ell}=30$~K with a
constant-volume, constant-temperature (NVT) run, $10$~ns long.  
A second, higher temperature value, $T_h=50$~K, is also considered to match the experimental conditions,
where both $T_{\ell}$ and $T_h$ are considered. The initial configurations  at $T_h$ are  generated from those 
at the end of the $T_{\ell}$ runs,  by increasing sharply 
the temperature and letting the systems equilibrate for further $10$~ns. 
The temperature is controlled by means of Langevin thermostats,\cite{langevin} with 
two separate thermostats used for the Xe/GL system, one for each atomic species,  
and only one for the Xe/Au system. The same integration time step $\delta t =1$~fs is used in all calculations.

The MD simulations are started from the initial conditions constructed by assigning to
the xenon atoms random, non-overlapping, positions on the surface, and random velocities
sampled from the Maxwell distribution at the temperature $T_o=25$~.
The system is then thermalized at $T_{\ell}=30$~K with a
constant-volume, constant-temperature (NVT) run, $10$~ns long.
A second, higher temperature value, $T_h=50$~K, is also considered to match the experimental conditions,
where both $T_{\ell}$ and $T_h$ are considered. The initial configurations at $T_h$ are generated from those
at the end of the $T_{\ell}$ runs, by increasing sharply
the temperature and letting the systems equilibrate for further $10$~ns.
The temperature is controlled by means of Langevin thermostats~\cite{langevin}, with
two separate thermostats used for the Xe/GL system, one for each atomic species,
and only one for the Xe/Au system. The same integration time step $\delta t =1$~fs is used in all calculations.

\section{Results \& discussion}

\subsection{Sampling the PES for xenon on graphene and gold by \emph{ab initio} calculations} \label{subsec:PES}

\begin{figure*}[htpb] 
\begin{center}
\includegraphics[width=0.48\textwidth]{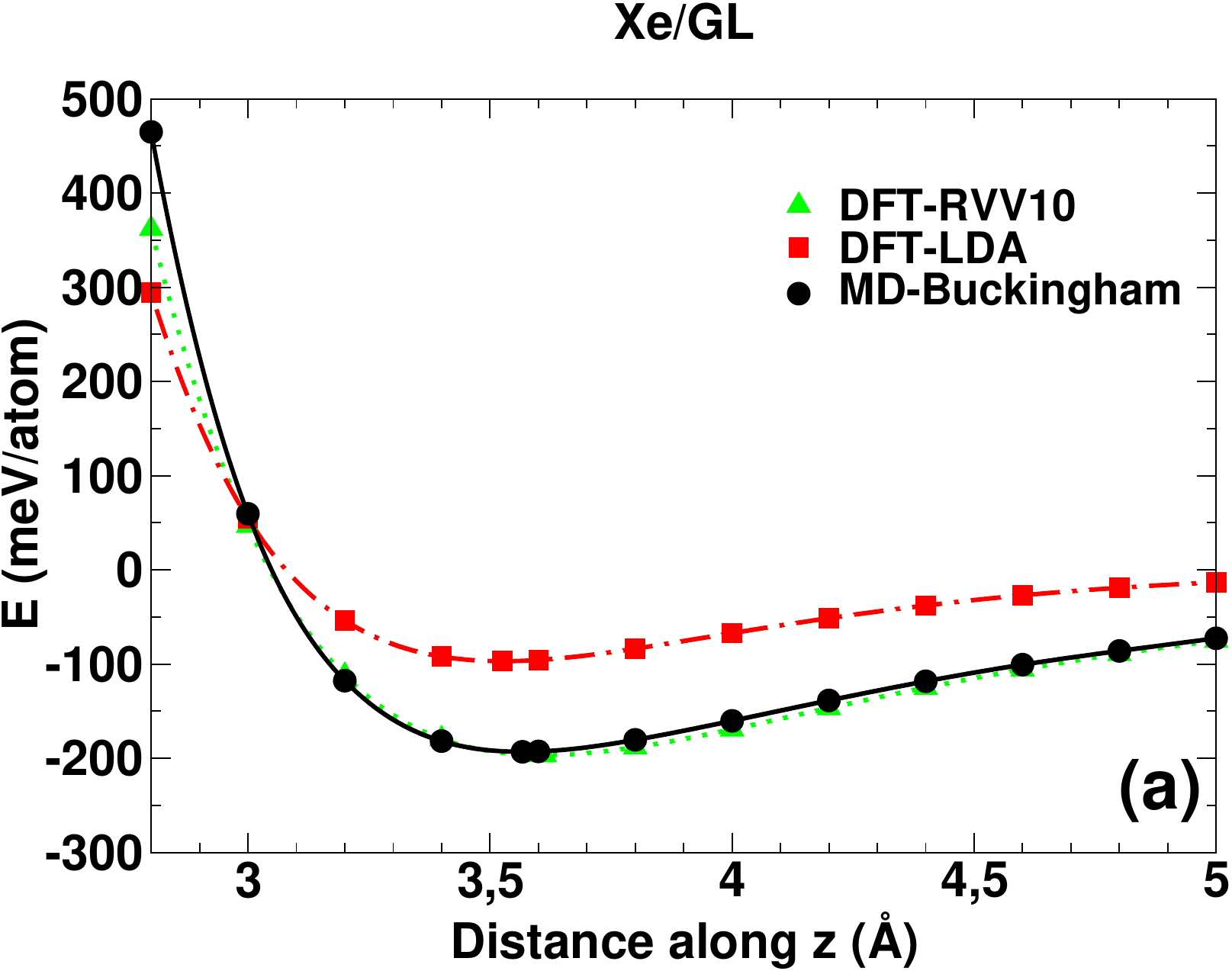}
\includegraphics[width=0.48\textwidth]{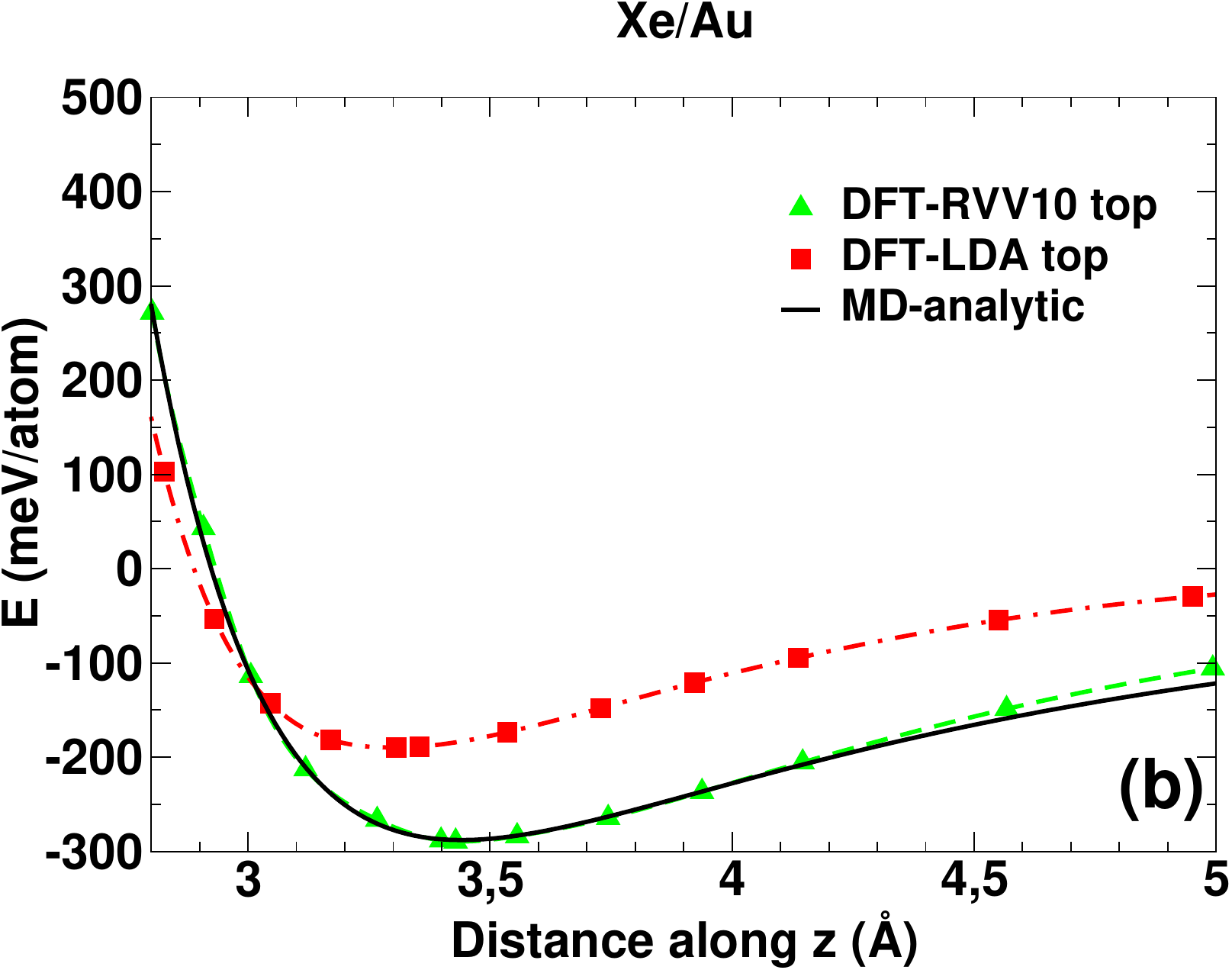}
\includegraphics[width=0.48\textwidth]{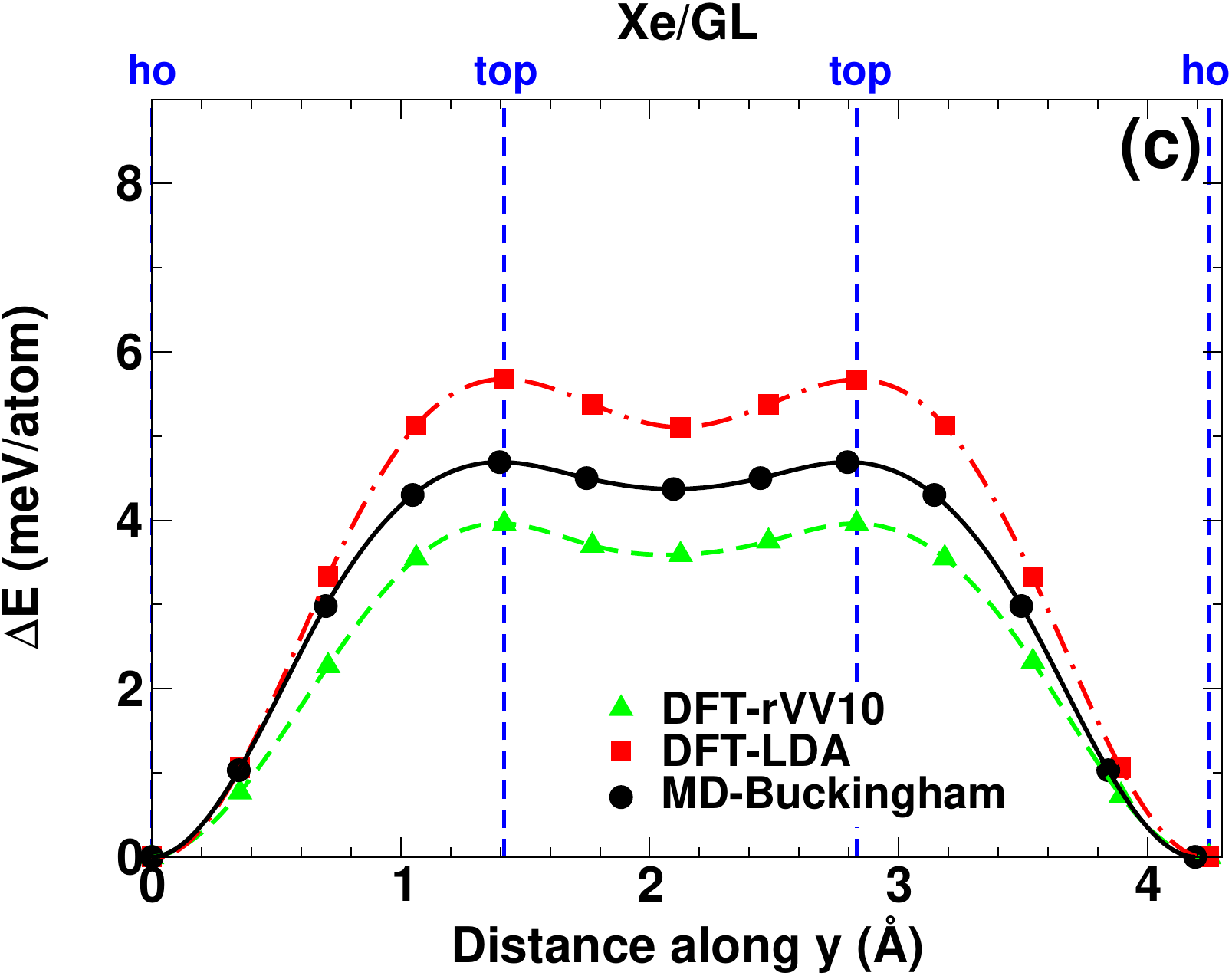}
\includegraphics[width=0.48\textwidth]{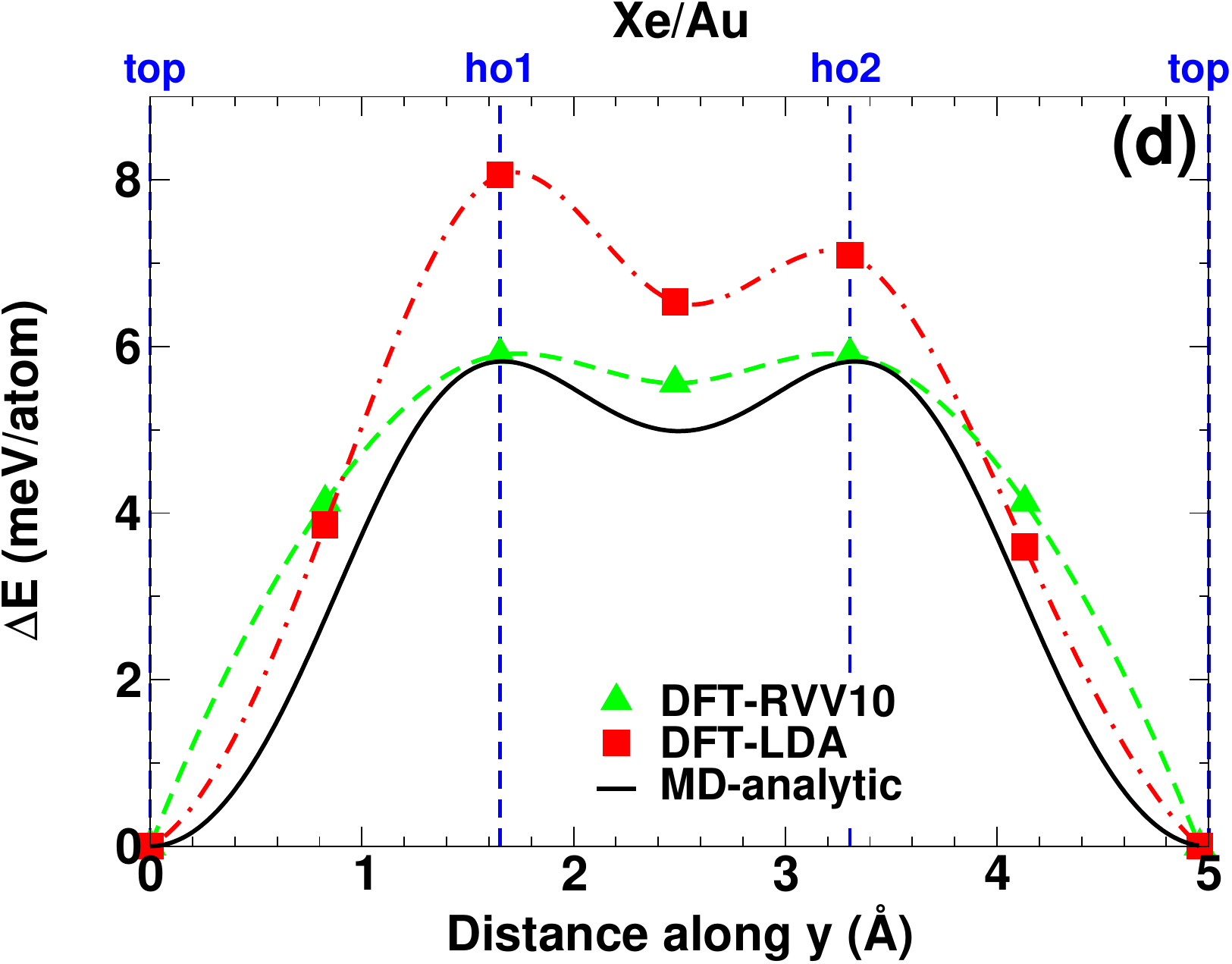}
\end{center}
\caption{Interaction energy between Xe and graphene (gold) as a function of the separation (a and b) and the relative lateral position (c and d). The latter spans the high symmetry sites labeled according to Fig.~\ref{fig:xe}.}
\label{fig:gpes}
\end{figure*}

We calculate the Xe adsorption energy on graphene and gold substrates
for different relative positions, obtaining in this way a sampling of the PESes
for the two adsorbate systems shown in Fig.~\ref{fig:xe}.
We firstly perform standard DFT calculations with the exchange correlation functional described by LDA.
Then, we take into account the vdW interactions by means of the rVV10 method~\cite{rvv10},
as motivated in the previous section. The results obtained within rVV10 are used as a benchmark
to derive the parameter values for the xenon–carbon (Eq.~\ref{eq:buck}) and xenon–Au(111) (Eq.~\ref{eq:potxe_onmetal})
potentials used in the MD simulations.

The calculated interaction energy, $E$, between the Xe adatom
and the substrate (see Section~\ref{sec:method} for the definition of $E$)
is represented as a function of the adatom–substrate separation, $z$,
in Fig.~\ref{fig:gpes} and b. The Xe atom position relative to the substrate
corresponds to the most favorable one in both the cases.
By comparing the red and green curves, it appears evident that the
DFT-LDA description underestimates the depth of the potential
wells, a problem which is solved by the rVV10 method that
increases the strength and widens the range of the attractive
part of the interactions.
The equilibrium distance, $z_{eq}$, and the corresponding energy in the minimum,
i.e., the adsorption energy $E_{ads}$, are reported in Table~\ref{tab:corr} both for
the Xe/GL and Xe/Au systems. It can be seen that the absolute value of the
adsorption energy of Xe on the gold substrate is significantly
higher, about 100 meV per atom, compared with that on the
graphene substrate; this is most likely due to the partial
hybridization of Xe 5p orbitals with Au d states~\cite{Zhang2014a,Zhang2014b}.
The equilibrium adsorption energies and distances are in good agreement
with previous experimental and theoretical data. In particular, the
Xe adsorption energy on graphene, $E^{\text{GL}}_{\text{ads}} = - 197$~meV per atom,
is within the range of theoretical data available in the literature
(–128.6~meV per atom by DFT/vdW-WF~\cite{Ambrosetti2011}, –142.9~meV per atom by MP2~\cite{mp2},
–204~meV per atom by all-electron full-potential linearized augmented plane wave
plus local orbitals method~\cite{DaSilva2007} and –209.7~meV per atom by the DFT/vdW-DF method~\cite{Zhang2014b}),
and the adsorption distance, $z^{\text{GL}}_{\text{eq}} = 3.61$~\r{A}, is in excellent agreement with
the experimental value of $3.59 \pm 0.04$~\r{A}. In the case of the Xe/Au system,
our adsorption energy, $E^{\text{Au}}_{\text{ads}} = -289$~meV per atom, is
in good agreement with previous theoretical calculations (–262.3~meV per atom by DFT/vdW-DF~\cite{Zhang2014b})
and within the experimental range~\cite{nieuwenhuys,Springborg2006,Forster2006,Andreev2004},
while the equilibrium distance, $z^{\text{Au}}_{\text{eq}} = 3.43$~\r{A},
is slightly shorter than the experimental (3.62~\r{A}) and theoretical (3.58~\r{A})
values reported in the literature~\cite{nieuwenhuys,Zhang2014b}.

\begin{table*}[htpb]
\caption{Optimized adsorption distance,  $z_{eq}$ (\r{A}), adsorption energy, $E_{ads}$ (meV per atom), and potential corrugation, $\Delta E$ (meV), for xenon on graphene and gold. The DFT results obtained within the LDA and rVV10 approaches are compared to other theoretical and experimental data present in the literature.} \label{tab:corr}
\begin{center}
\begin{ruledtabular}
\begin{tabular}{c c c c c c c}
      &$z_{eq}^{GL}$ &$z_{eq}^{Au}$  &$E_{ads}^{GL}$ &$E_{ads}^{Au}$ & $\Delta E^{GL}$ & $\Delta E^{Au}$\\
\hline
DFT-LDA              &  $3.53$  & $3.31$ &  $-97$  & $-190$ & $6$  & $8$  			    \\
DFT-rVV10            &  $3.61$  & $3.43$ &  $-197$ & $-289$ & $4$  & $6$		 	    \\
Theory          &  $3.56 \div 4.18^{\text{a,b,c}}$  &  $3.58^{\text{a}}$   &  $-128.6 \div -209.7^{\text{a,b,c,d}}$ &  $-262.3^{\text{a}}$     & $3.1 \div 15.5^{\text{a,c,e}}$ &  $2.7^{\text{a}}$                 \\
Experiments         &  $3.59 \pm 0.04^{\text{f}}$ & $3.62^{\text{g}}$ & $-$ & $-199.5 \div -363^{\text{g,h,i,j}}$ & $3.29 \div 5.3^{\text{k,l,m}}$ & $2.2^{\text{l}}$\\
\end{tabular}
\end{ruledtabular}
\end{center}
\begin{flushleft}
{\footnotesize
$^{\text{a}}$Ref.~\onlinecite{Zhang2014b},
$^{\text{b}}$Ref.~\onlinecite{Ambrosetti2011},
$^{\text{c}}$Ref.~\onlinecite{mp2},
$^{\text{d}}$Ref.~\onlinecite{DaSilva2007},
$^{\text{e}}$Ref.~\onlinecite{steele},
$^{\text{f}}$Ref.~\onlinecite{pussi} for Xe/graphite,
$^{\text{g}}$Ref.~\onlinecite{nieuwenhuys},
$^{\text{h}}$Ref.~\onlinecite{Springborg2006},
$^{\text{i}}$Ref.~\onlinecite{Forster2006},
$^{\text{j}}$Ref.~\onlinecite{Andreev2004},
$^{\text{k}}$Ref.~\onlinecite{Coffey2005},
$^{\text{l}}$Ref.~\onlinecite{mistura},
$^{\text{m}}$Ref.~\onlinecite{hong} for Xe/graphite
}
\end{flushleft}
\end{table*}

We then calculate the Xe–substrate interaction energy for different relative lateral positions.
In particular, we consider the configurations along the direction (referred to as the $y$ direction)
passing through the high symmetry sites labelled in Fig.~\ref{fig:xe}. The Xe–substrate distance is optimized
at each location. The energy profile obtained for the Xe/GL system, represented in Fig.~\ref{fig:gpes}c,
presents the minima (maxima) at hollow (on-top) sites. In contrast, the energy profile obtained for the Xe/Au system
presents the minima (maxima) at on-top (hollow) sites, in agreement with experimental observations.
The energy difference between the maxima and the minima ($\Delta E$), referred to as the PES corrugation,
is reported in Table~\ref{tab:corr}. The PES corrugation calculated within both the DFT schemes is slightly
higher for the Xe/Au system than for Xe/GL. The order of magnitude of the PES
corrugations is the same as those estimated from experiments.
This is consistent with the fact that Xe presents a stronger binding on gold
than on graphene. It is, in fact, often observed that the energy barriers to
displace an adsorbate (or a countersurface) along a substrate increase with
the strength of the adsorbate–substrate interaction.

\begin{table}[htpb]
\caption{Values of the $A_B, b_B, C_B$ parameters for the Xe–C interaction and of 
the $ A_0,A_1, A_2$ coefficients in the functions $A_i(x,y)$ for the  Xe–Au(111) interaction.}\label{tab:aicoeff}
\begin{center}
\begin{ruledtabular}
\begin{tabular}{c c c c}
&&Xe-C &  \\ 
 Buckingham   & $A_B $ (eV) & $b_B  (\text{\r{A}}^{-1}) $ & $C_B (\text{eV} \cdot \text{\r{A}}^{6})$\\
 			  & $1.835 \cdot 10^{4}$   &         $0.273$     		  & $76.494$ \\
\hline
&&Xe-Au(111)  &\\
Analytic & $A_0$ (eV) & $A_1$ (\r{A}) & $A_2$ ($\text{eV} \cdot \text{\r{A}}^3$) \\
\hline
$top$                   &  $3.88 \cdot 10^{4}$  &  $3.78$    &  $15.24$                \\
$ho$                    &  $3.13 \cdot 10^{3}$  &  $2.99$    &  $15.79$                 \\
\end{tabular}
\end{ruledtabular}
\end{center}
\end{table}

The results obtained within the DFT-rVV10 scheme for the adsorption of Xe on both graphene and gold substrates
appear to be accurate enough to be used as the reference
dataset for tuning the unknown parameters appearing in the force field for the Xe–C (Eq.~\eqref{eq:buck})
and Xe–Au (Eq.~\eqref{eq:potxe_onmetal}) interactions.
By fitting the data shown in Fig. \ref{fig:gpes}a and b using a nonlinear least-squares
Marquardt–Levenberg algorithm~\cite{gnuplot}, we obtain the parameter
values reported in Table~\ref{tab:aicoeff}.
The fitting is very satisfactory as can be seen in Fig.~\ref{fig:gpes}a and b by comparing
the Xe adsorption energy as a function of the adatom–substrate separation
calculated within the MD scheme (in black color) with the
\textit{ab initio} rVV10 data (in green color); in particular the equilibrium
energies and distances are in excellent agreement.
The energy variation as a function of the adatom lateral position,
shown in Fig.~\ref{fig:gpes}c and d, is well reproduced by the adopted force fields.
An accurate description of the lateral PES is essential for a reliable
simulation of tribological systems. The shape and the corrugation
of the PES determine, in fact, the frictional forces. It is typically
difficult to correctly reproduce the shape of the PES for rare gases
on metals by analytical force fields. Pair wise potentials, like the
Lennard-Jones potential, favor, in fact, the sites with the highest
coordination. Experiments reveal, instead, that for many rare
gas atoms on metals, such as Xe on gold, the adsorption occurs
on-top. Such hard substrates can be accurately modeled by an
external potential representing the interaction with a fixed and
rigid triangular lattice frame~\cite{Varini2015,modelxe}.
The potential we adopted to model the Xe/Au system correctly reproduces
the PES shape (Fig.~\ref{fig:gpes}), with minima at on-top sites and maxima at hollow
sites. Not only the potential corrugation is in excellent agreement
with that obtained by the rVV10 method. A small underestimation,
less than 1 meV, is instead observed in the PES corrugation for
the Xe/GL system. The uncertainty in the estimation of the PES
corrugation is not expected to affect the results of the MD
simulations presented in the next section. The different lattice
match of the Xe/Au and Xe/GL plays, in fact, a major role in
determining the degree of commensurability of the rare gas
islands in the two systems.

In conclusion, the parametric potentials tuned on the
DFT-rVV10 data provide an accurate description of the adsorbate systems. Their use in MD simulations
will allow us to study the evolution of a system relevant for tribology, containing a large
number of xenon atoms over long time scales, typically inaccessible by \textit{ab initio} MD~\cite{cp,my_prl13}.

\subsection{Monitoring the commensurability of the adsorbed monolayers by molecular dynamics }
\label{MD}

\begin{figure*}[htpb]
\begin{center}
\begin{subfigure}{\linewidth}
\Large{\textbf{Xe/GL}} \hspace{6.5cm} \Large{\textbf{Xe/Au}}
\end{subfigure}
\begin{subfigure}{0.24\linewidth}
\includegraphics[width=0.9\textwidth]{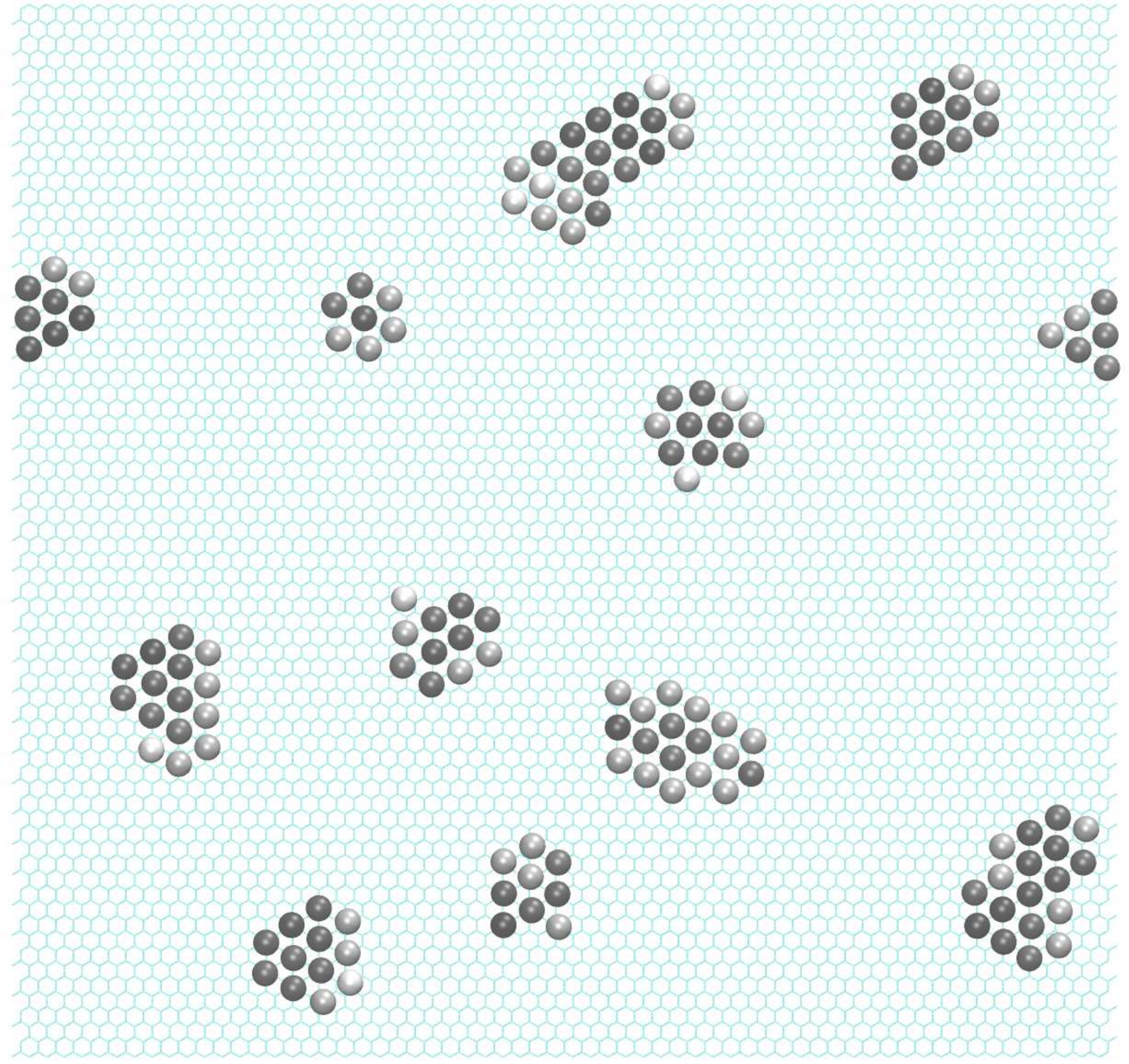}
\caption{$\theta=11\%\,,\; T=30$ K}
\label{colorgl_1030}
\end{subfigure}
\begin{subfigure}{0.24\linewidth}
\includegraphics[width=0.9\textwidth]{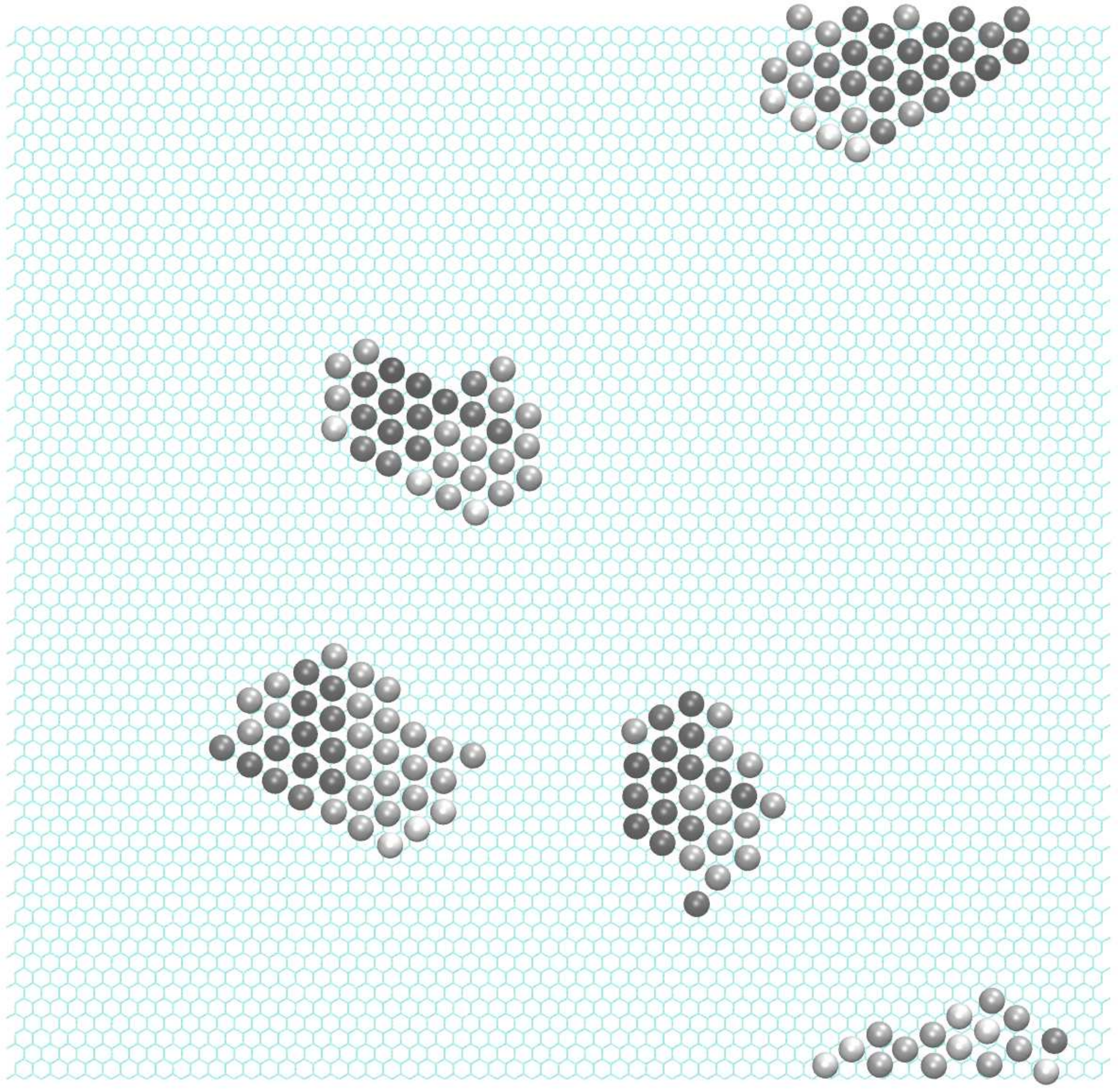}
\caption{$\theta=11\%\,,\; T=50$ K}
\label{colorgl_1050}
\end{subfigure}
\begin{subfigure}{0.24\linewidth}
\includegraphics[width=0.9\textwidth]{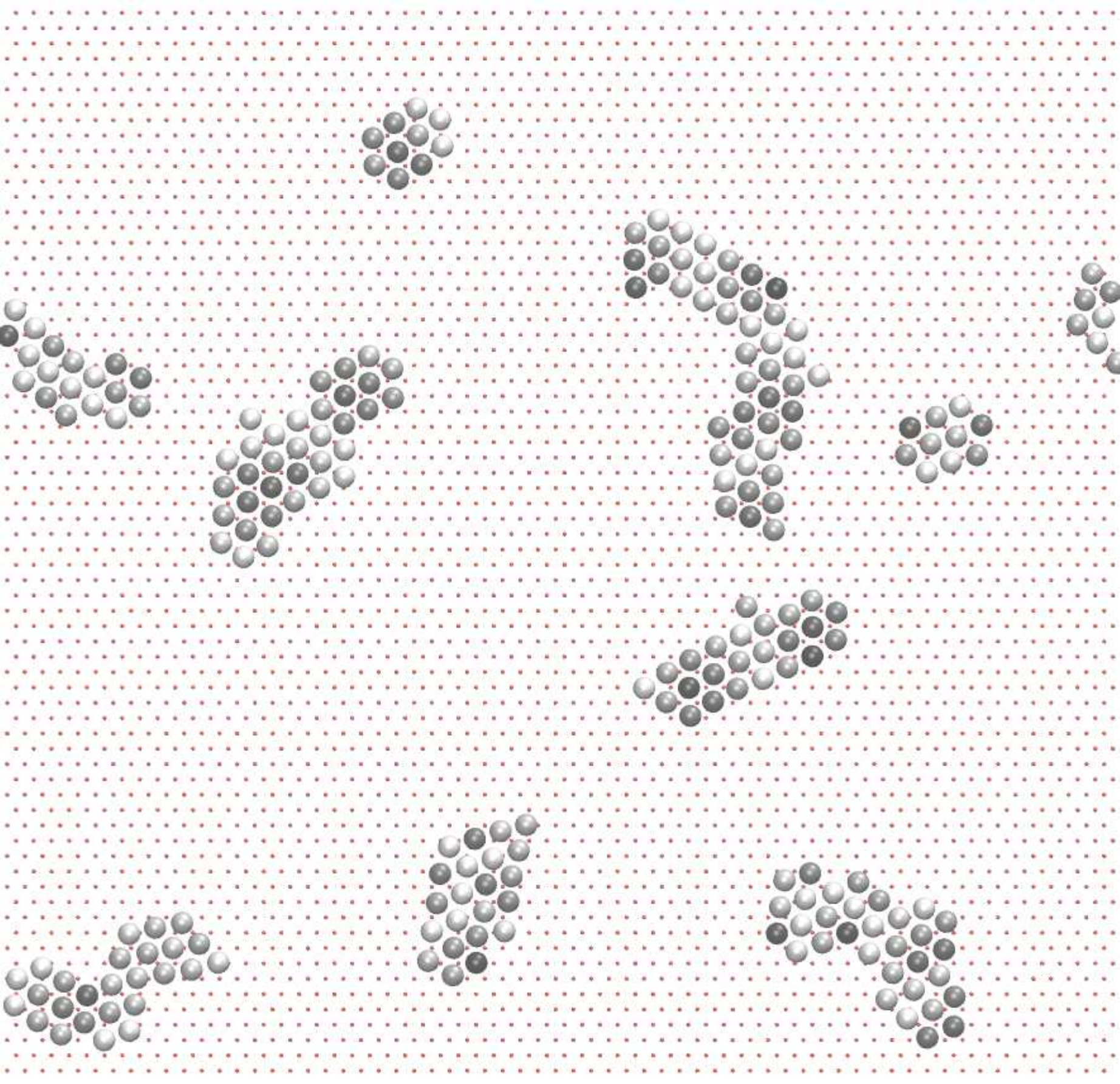}
\caption{$\theta=11\%\,,\; T=30$ K}
\label{colorau_1030}
\end{subfigure}
\begin{subfigure}{0.24\linewidth}
\includegraphics[width=0.9\textwidth]{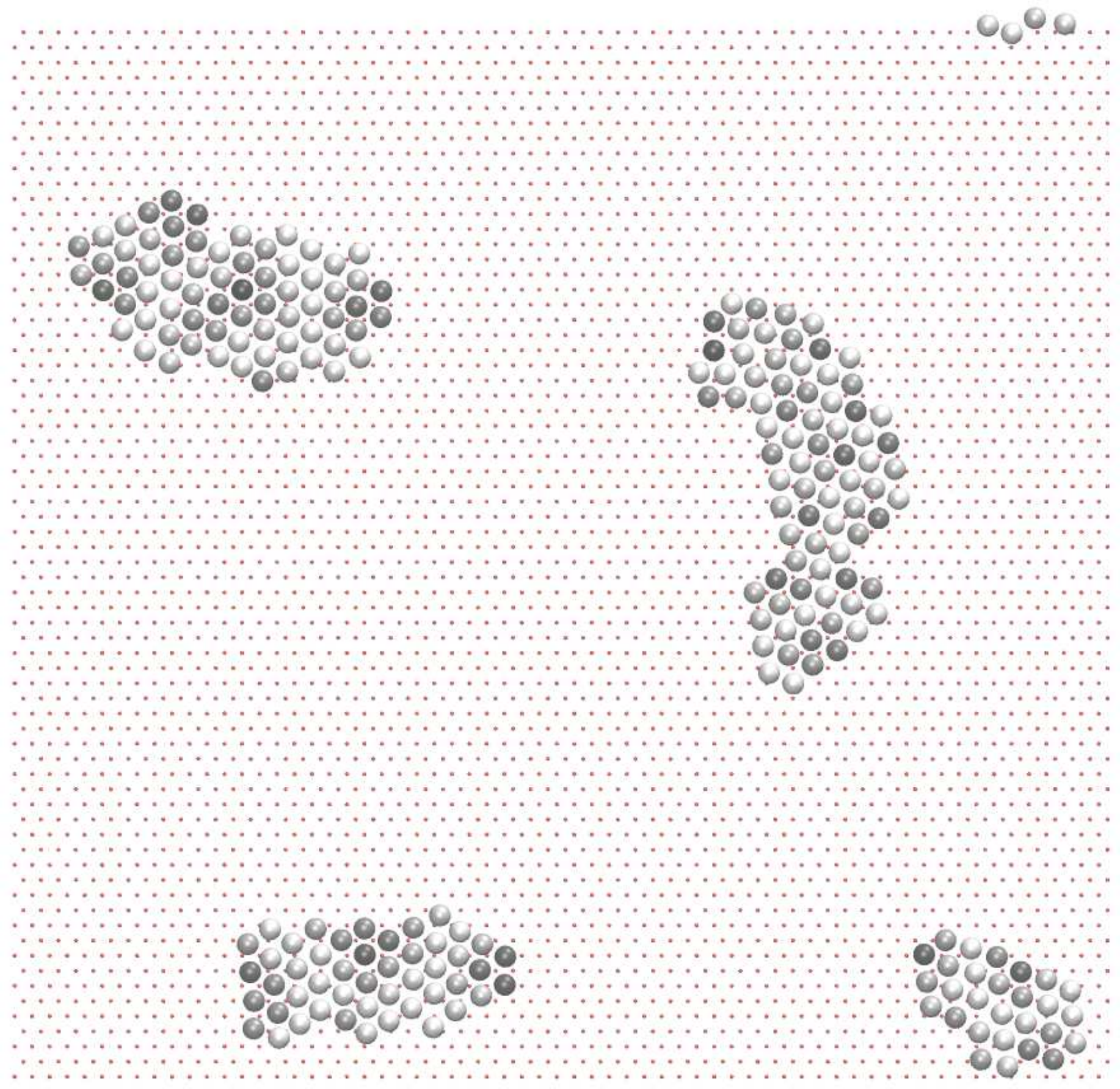}
\caption{$\theta=11\%\,,\; T=50$ K}
\label{colorau_1050}
\end{subfigure}

\begin{subfigure}[b]{0.24\linewidth}
\includegraphics[width=0.9\textwidth]{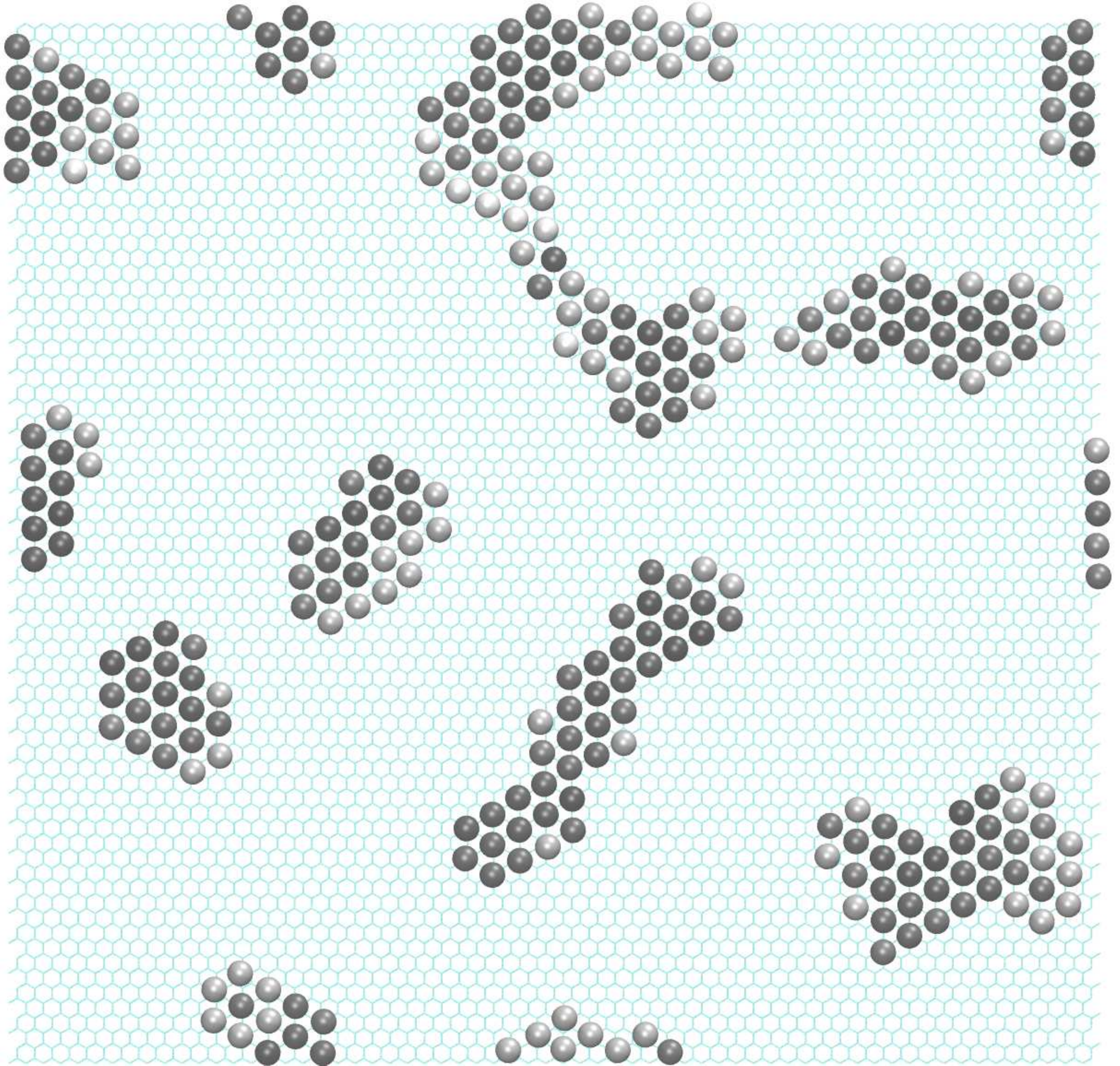}
\caption{$\theta=22\%\,,\; T=30$ K}
\label{colorgl_2030}
\end{subfigure}
\begin{subfigure}[b]{0.24\linewidth}
\includegraphics[width=0.9\textwidth]{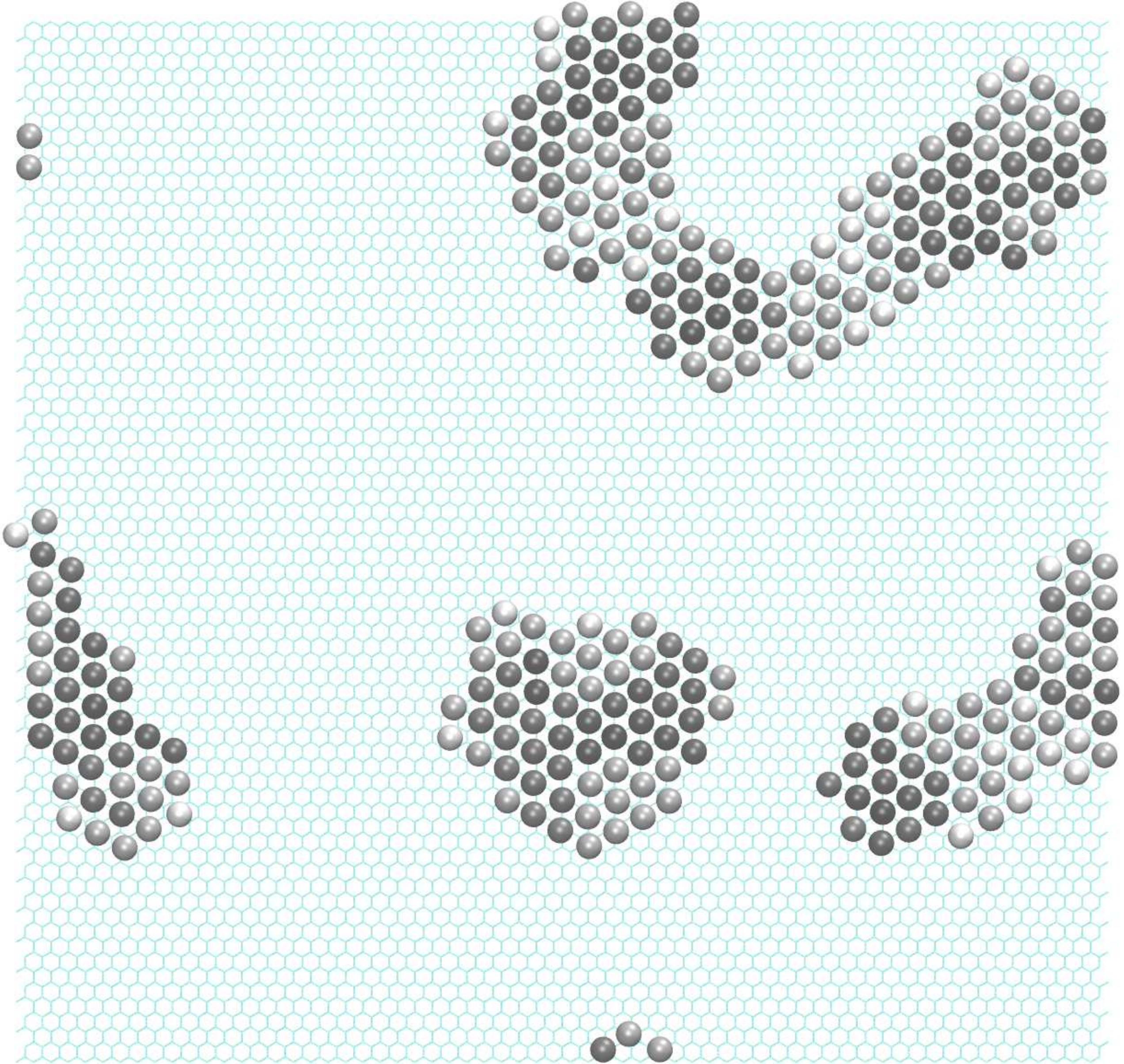}
\caption{$\theta=22\%\,,\; T=50$ K}
\label{colorgl_2050}
\end{subfigure}
\begin{subfigure}[b]{0.24\linewidth}
\includegraphics[width=0.9\textwidth]{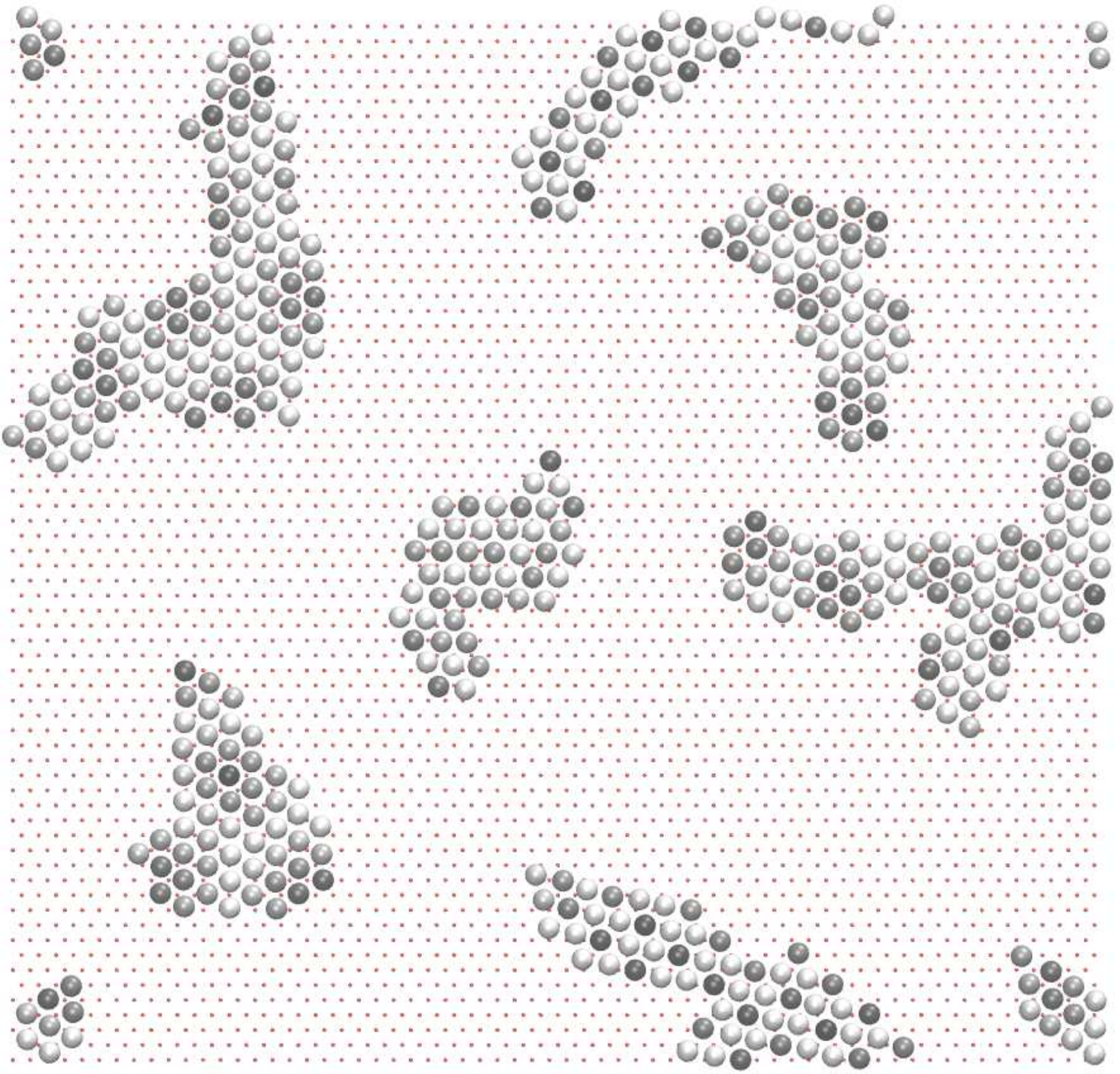}
\caption{$\theta=22\%\,,\; T=30$ K}
\label{colorau_2030}
\end{subfigure}
\begin{subfigure}[b]{0.24\linewidth}
\includegraphics[width=0.9\textwidth]{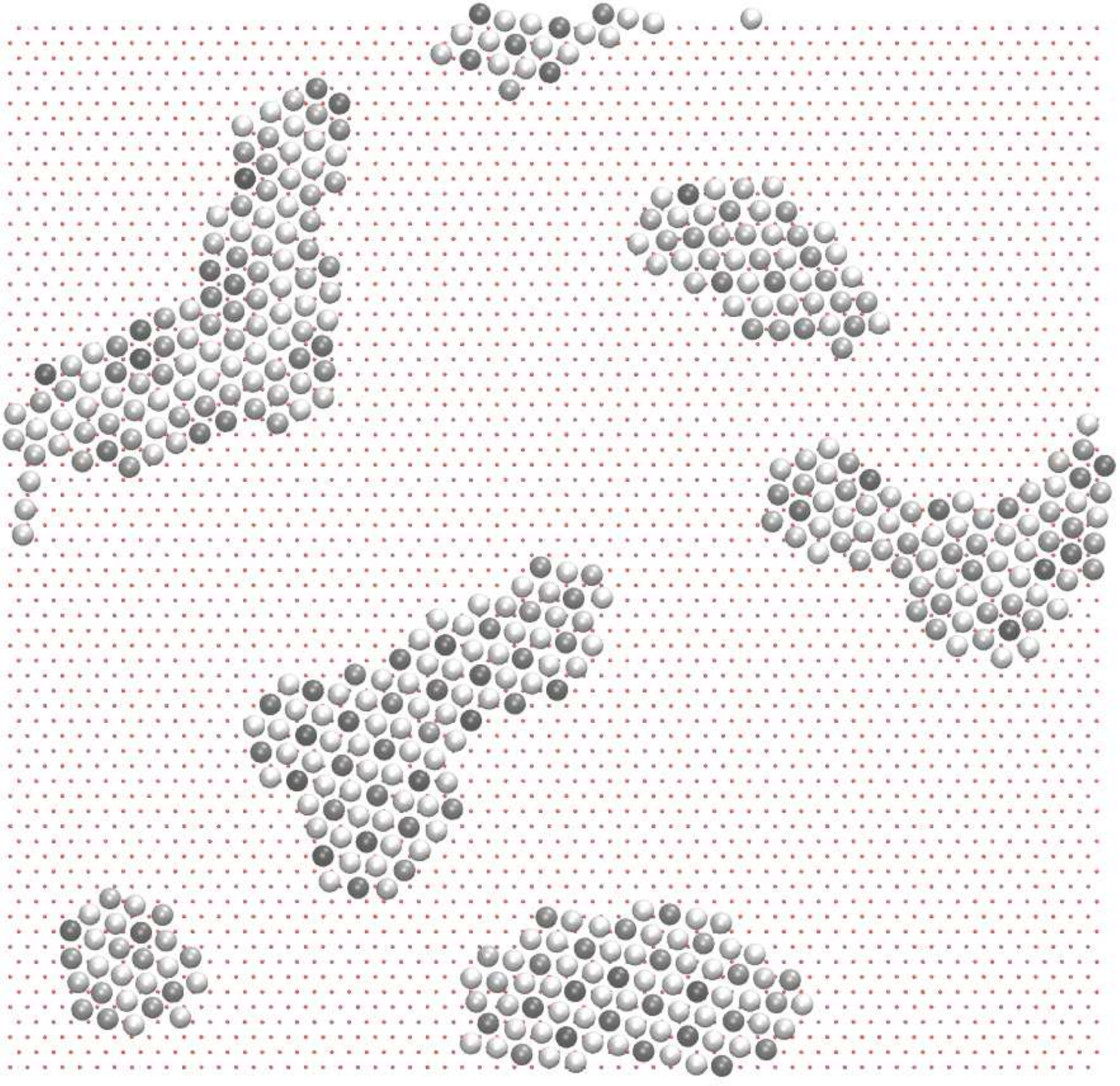}
\caption{$\theta=22\%\,,\; T=50$ K}
\label{colorau_2050}
\end{subfigure}

\begin{subfigure}[b]{0.24\linewidth}
\includegraphics[width=0.9\textwidth]{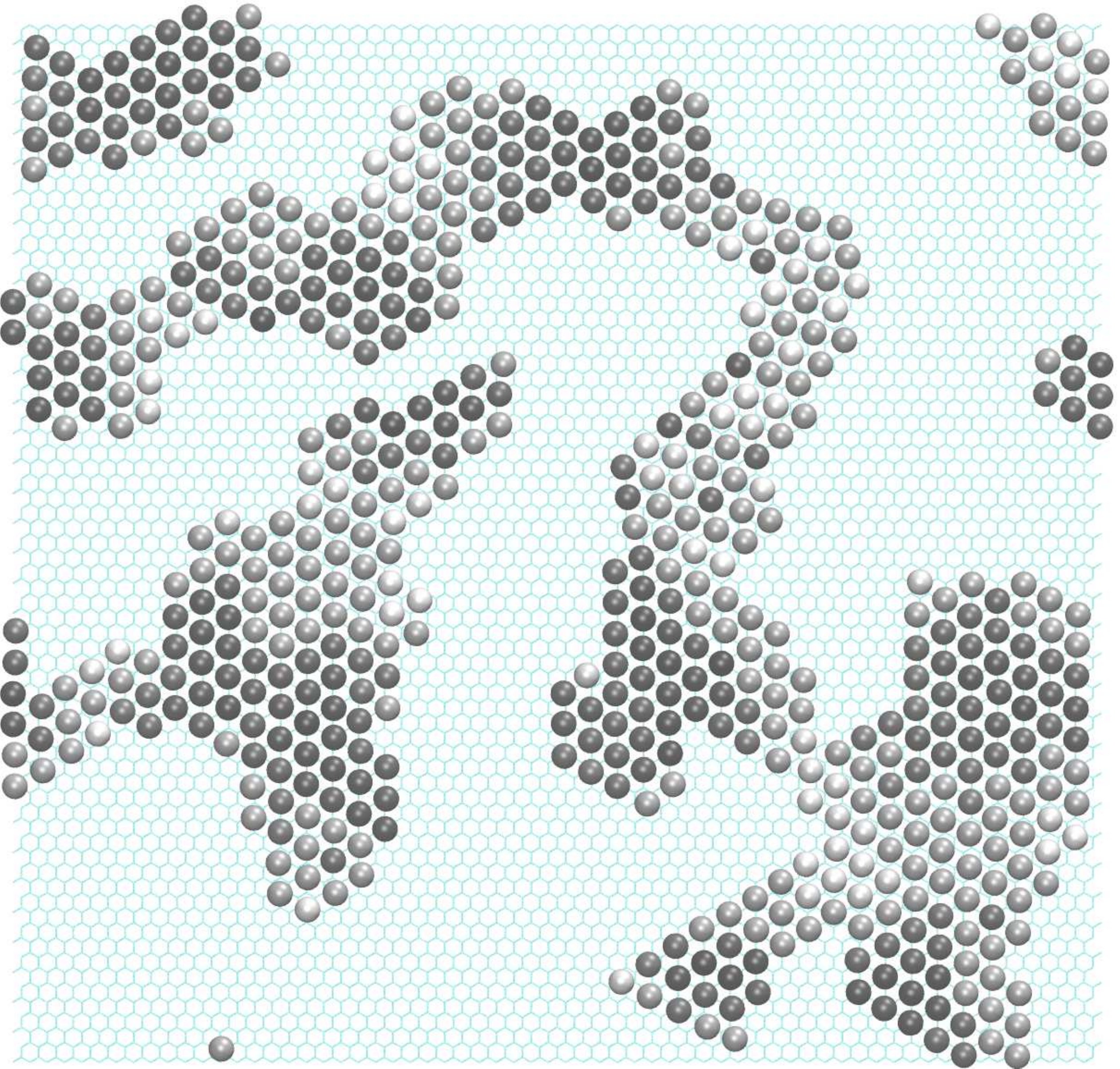}
\caption{$\theta=44\%\,,\; T=30$ K}
\label{colorgl_4030}
\end{subfigure}
\begin{subfigure}[b]{0.24\linewidth}
\includegraphics[width=0.9\textwidth]{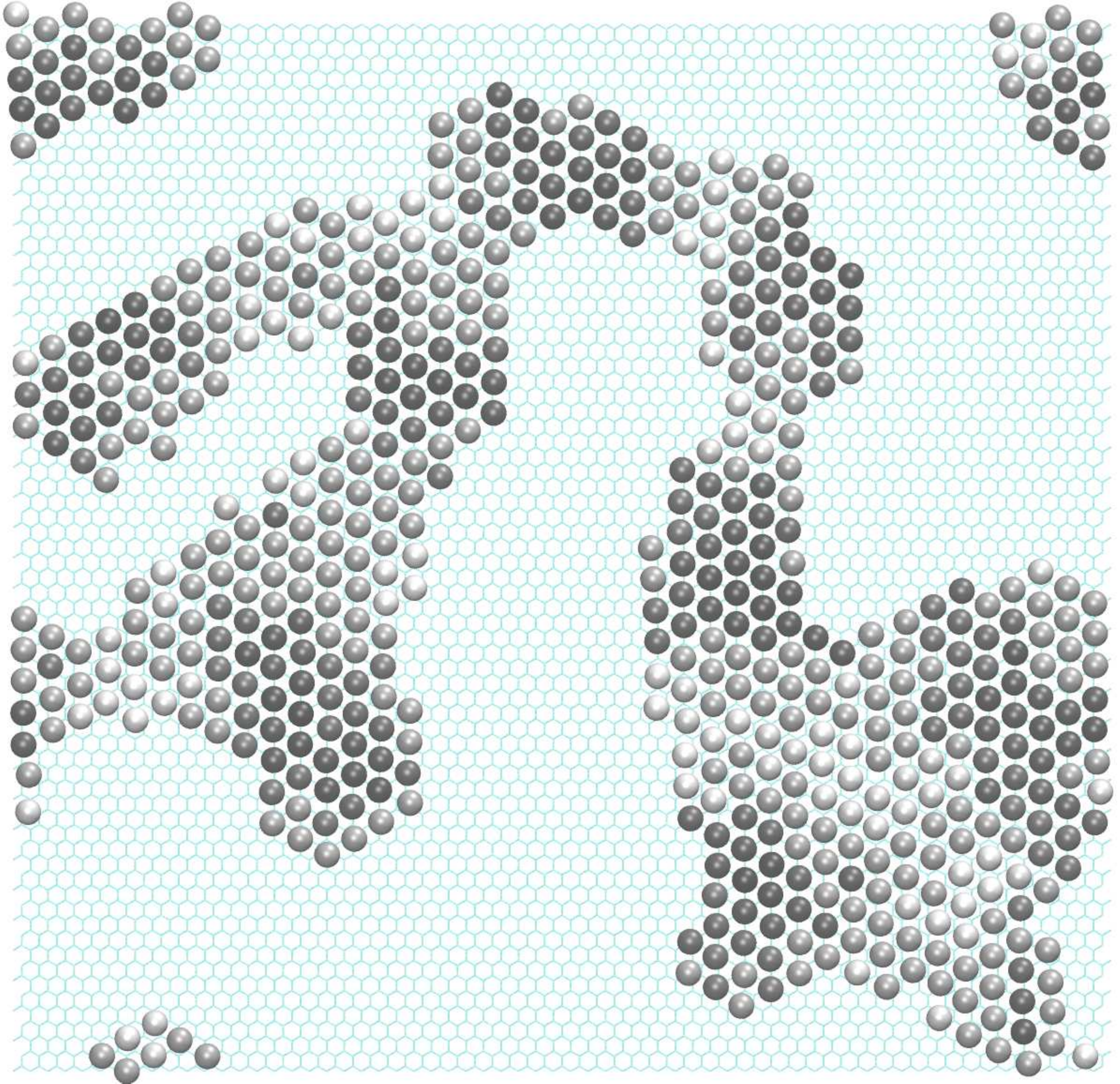}
\caption{$\theta=44\%\,,\; T=50$ K}
\label{colorgl_4050}
\end{subfigure}
\begin{subfigure}[b]{0.24\linewidth}
\includegraphics[width=0.9\textwidth]{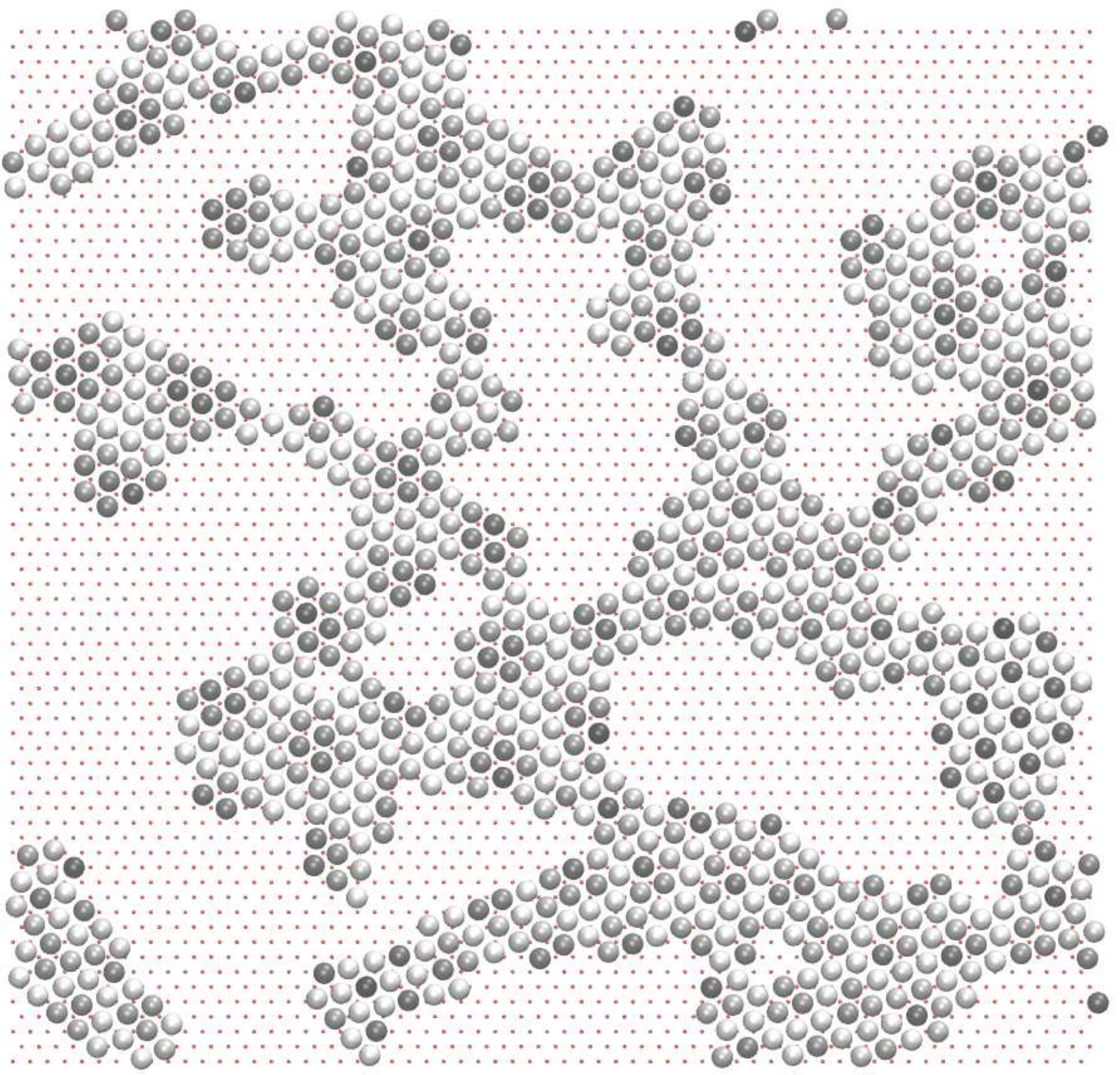}
\caption{$\theta=44\%\,,\; T=30$ K}
\label{colorau_4030}
\end{subfigure}
\begin{subfigure}[b]{0.24\linewidth}
\includegraphics[width=0.9\textwidth]{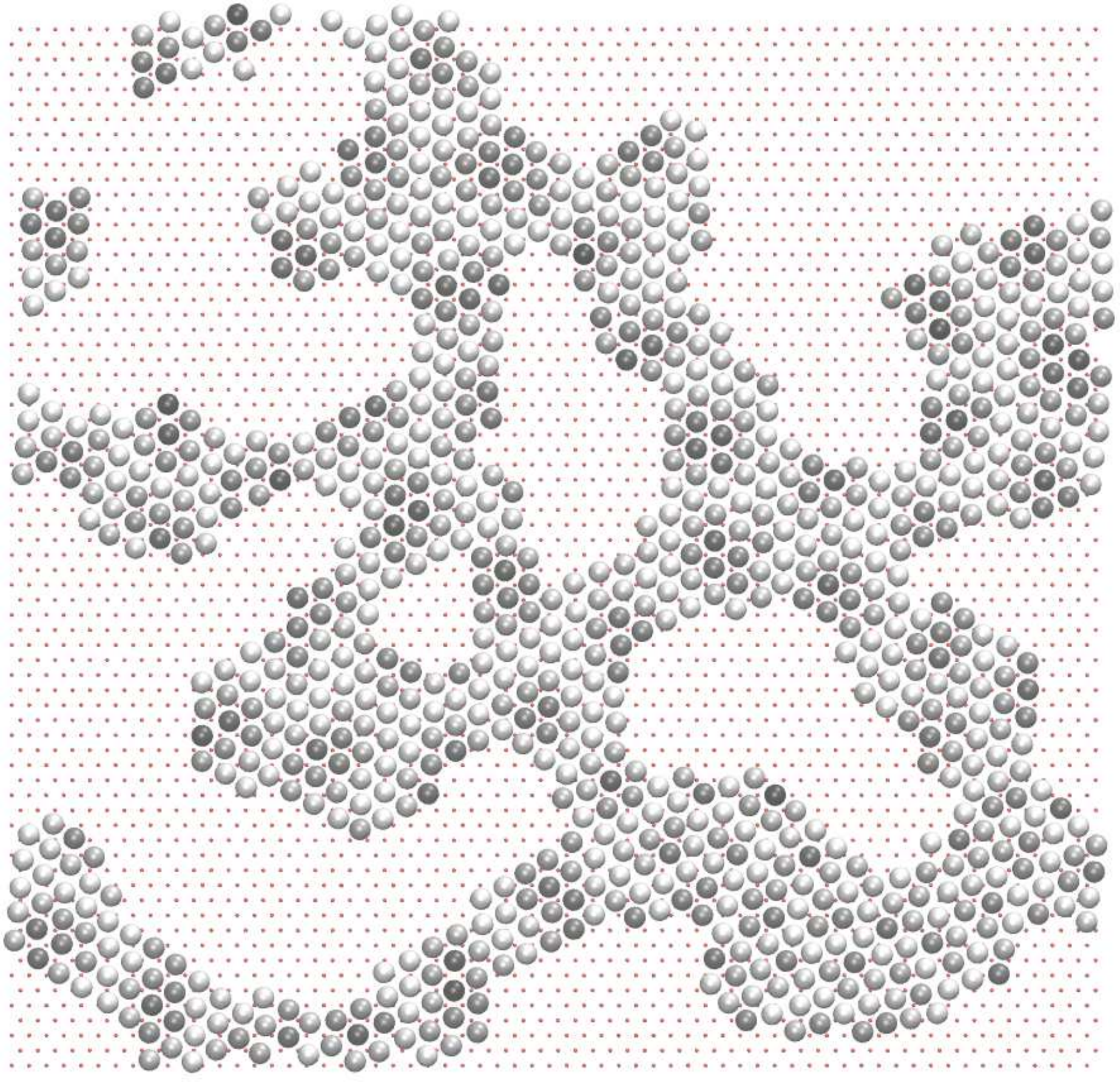}
\caption{$\theta=44\%\,,\; T=50$ K}
\label{colorau_4050}
\end{subfigure}
\end{center}
\caption{Snapshots of the simulation cell for the Xe/GL and Xe/Au systems at the end of MD simulations each lasted 20 ns. The simulated coverage, $\theta$, and temperature, $T$, are indicated in each figure panel. The Xe particles are colored according to their position on the PES: a grayscale is used to indicate the
particle distance from the PES minima (maxima), in black (white).
}\label{fig:colorxe}
\end{figure*}

Classical molecular dynamics simulation runs, each 20 ns long, for three coverage values and
two different temperatures have been analyzed in parallel for the Xe/GL and Xe/Au systems. The
final atomic configurations of all the simulated systems are pictured in Fig.~\ref{fig:colorxe}
using a range of gray shades from black to white to illustrate the different level of
commensurability between the xenon layer and the substrates below: black color means that the
adatom is commensurate with respect to the substrate (\textit{i.e.} resides in a minimum of the potential);
white color means that the adatom is fully incommensurate (\textit{i.e.} is located in a maximum of the
potential), and gray levels represent intermediate conditions.

In both the studied systems, it is possible to notice that the xenon atoms at low temperature
and low coverage gather into many separated islands of small size with the characteristic
hexagonal pattern. In the Xe/GL case there is a predominance of black as the Xe atoms tend to
occupy the PES minima; in this case the mismatch of the lattices is small and the number of
neighbors of each Xe atom is not enough to provide the energy gain to promote incommensurability
between the Xe and GL lattices. For the Xe/Au system, instead, while the size of the island is
comparable, the presence of black Xe atoms is much less relevant. 

With increasing temperature, one can see immediately that the size of these islands increases
in all cases, as the increase of thermal energy makes the Xe atoms to easily overcome the
energy barriers of the PES and Xe diffusion on the surface is correspondingly enhanced.
The grayscale patterns do not change significantly for the Xe/Au systems, but for the Xe/GL
systems one can immediately notice the presence of a lesser number of neighboring black Xe atoms.

As expected, increasing the coverage leads to an increase of the size of the clusters.
In particular, while at 11\% and 22\% coverages one can count several isolated ones, at 44\%
for both systems Xe atoms are arranged in a unique connected cluster, \textit{i.e.},
the size of the island reaches the dimension of the MD cell. In the Xe/Au case there are mainly
isolated black Xe atoms occupying the PES minima while for Xe/GL several, somewhat smaller,
black patches remain.

\begin{figure}[htpb] 
\centering
\includegraphics[width=0.48\linewidth]{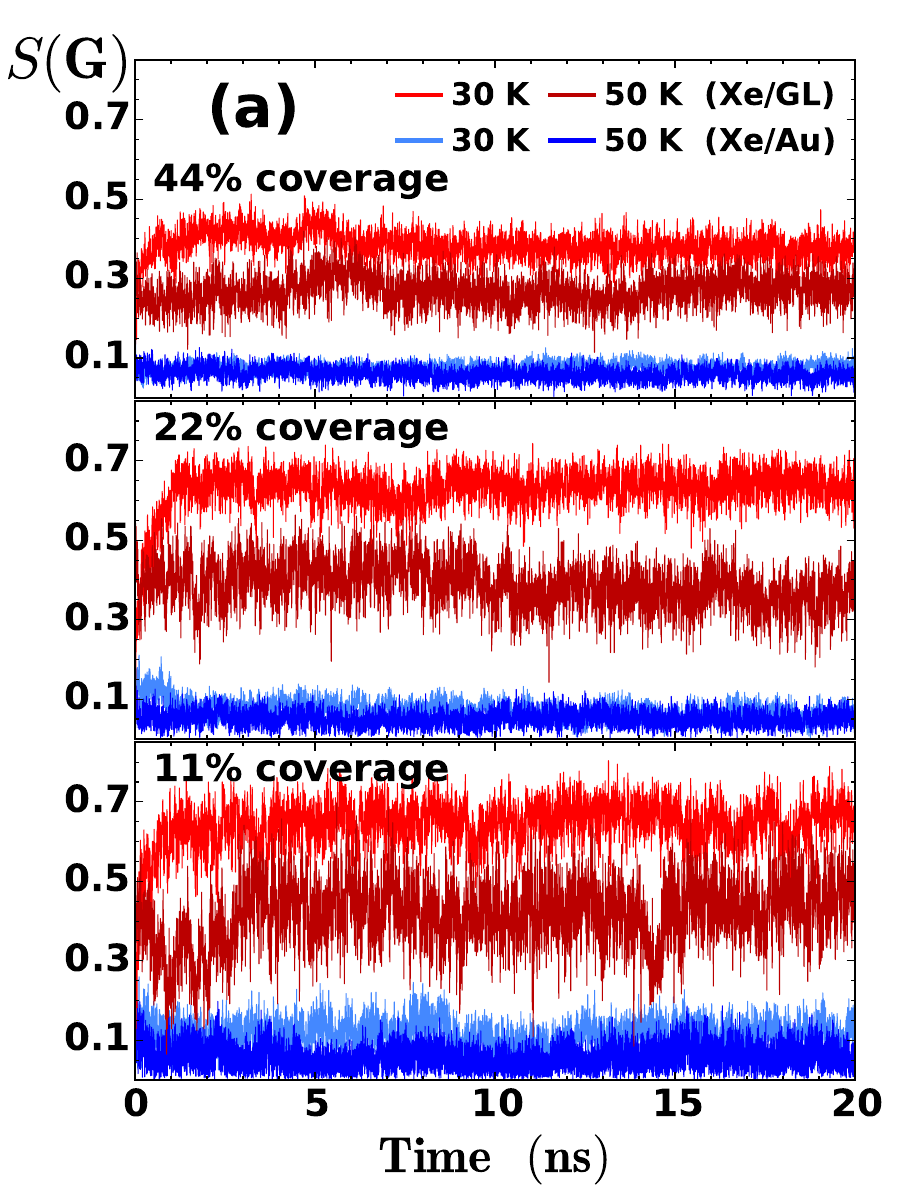}
\includegraphics[width=0.48\linewidth]{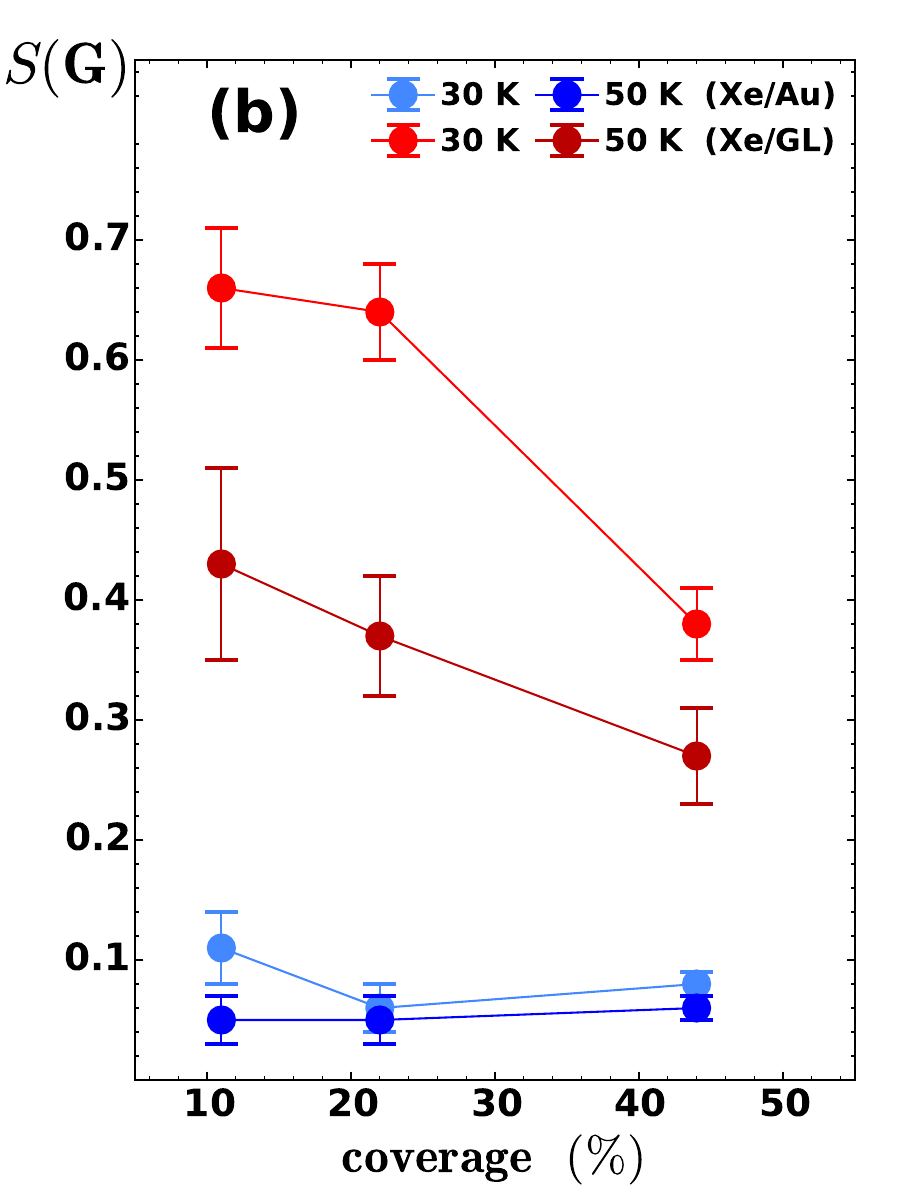}
\caption{The average structure factor $S(\mathrm{G})$ for the Xe/GL and Xe/Au systems (a) as a function of the time and
(b) evaluated averaging over the last 10 ns of the simulation. The two different scales of red (blue) represent the two
different temperatures used during the simulations for the Xe/GL (Xe/Au)
system. Reported error bars correspond to the standard deviations.}
\label{fig:sk_g}
\end{figure}

Overall one sees that the size of these xenon clusters increases with both coverage
and temperature and, correspondingly, their structure becomes more and more
incommensurate with the substrate. The change in the range of temperature and
coverage explored is more significant for the Xe/GL system, where the mismatch
between the Xe and the substrate is smaller, favoring the existence of small
commensurate islands at lower temperatures and coverages. This qualitative
behavior of the adatoms can be confirmed quantitatively by computing
the structure factor $S(\mathbf{G})$ of the xenon atoms

\begin{equation}
   S(\mathbf{G}) = \sqrt{\frac{\left\vert \sum_{j = 1}^{N} e^{i \mathbf{G} \cdot \mathbf{r}_j} \right\vert^2}{N^2}}
\end{equation}

\noindent where $\mathbf{G}$ is the summation of the two reciprocal lattice basis vectors
of the substrate cell  and $\mathbf{r}_{j}$ is the real space position of the $j$-th
xenon atom. $S(\mathbf{G})$ crosses over from $\sim 0$, for fully incommensurability,
to 1 for fully commensurability. In panel (a) of Fig.~\ref{fig:sk_g} we report the behaviors
of each individual $S(\mathbf{G})$ as a function of time, together with the averaged values of 
$S(\mathbf{G})$ over the last 10 ns of the simulations for both systems. It is
possible to notice that, by varying the coverage and the temperature of the systems,
$S(\mathbf{G})$ changes significantly; in particular, at low coverage and low temperature, 
$S(\mathbf{G})$ attains its maximum, instead at high coverage and temperature,
$S(\mathbf{G})$ decreases significantly in both systems. Moreover, $S(\mathbf{G})$
is always higher in the Xe/GL system than in the Xe/Au one. These data allow us to
confirm quantitatively the qualitative analysis based on the final atomic configurations.

Our results complemented with a previous study, where we described the close
relation between the commensurability and static friction of adsorbed islands~\cite{Reguzzoni2012},
can provide an explanation for the QCM experimental observations reported in Ref.~\cite{mistura}.
In particular, (i) the existence of pinning forces in nominally incommensurate systems at
low coverage can be accounted for by the result that small adsorbed islands are commensurate
with the substrate even in the presence of lattice misfit. (ii) The existence of a critical
coverage necessary for the film slipping is related to the critical size that the growing
Xe islands need to reach for the depinning process to be activated at the considered temperature.
(iii) Lower critical coverages are observed for Xe/Au than for Xe/GL in spite of the larger PES
corrugation of the former system because the lattice misfit has a larger impact on the commensurability
of islands. We showed, in fact, that the critical size for adsorbed islands,
$R_C \sim \frac{1}{e^2}\sqrt{\frac {\Delta E}{\epsilon}}$, depends on the ratio between the potential
corrugation, $\Delta E$, and the interparticle interaction strength $\epsilon$, and is dominated by
the lattice misfit, $e$~\cite{Reguzzoni2012}. We expect that the above result can be generalized to
solid clusters on crystalline substrates: clusters will interlock with substrates if their size
shrinks to a critical value. How small is this value depends on the bulk modulus, interfacial corrugation
and lattice mismatch. An example of this situation may be provided by AFM tips that quite often display a
stick-and-slip behavior, due to interlocking, independently of the nominal incommensurability of the tip and
the substrate materials.

\section{Conclusions}

In a recent experiment~\cite{mistura}, the nanofriction of Xe monolayers
deposited on graphene has been measured by means of a
quartz crystal microbalance. At low temperatures, the adsorbate
was fully pinned to the substrate, and it started sliding as soon
as a critical coverage was reached. The critical coverage was
found to depend on the temperature, in particular it decreases
with the temperature. Similar measurements repeated on bare
gold showed an enhanced slippage of the Xe films and a
decrease of the depinning temperature.

To shed light on the atomistic mechanism governing the
above described experimental observations, we have performed
a comparative study of xenon dynamics on graphene and gold
substrates at different coverages and temperatures. The first
part of our work is devoted to the accurate theoretical description
of the Xe–graphene and Xe–Au(111) interactions. Our DFT calculations show that the
rVV10 method~\cite{rvv10}, which allows to treat nonlocal van der Waals interactions
from first principles, improves the LDA description and provides adsorption energies and
distances in very good agreement with the available experimental data. The DFT rVV10 results
are used to optimize the force fields employed in the MD simulations. An excellent fit of the PES
for the Xe/GL system is obtained by using the Buckingham pairwise potential for the
Xe–C interactions, while the Xe–Au(111) interaction is accurately described by means of
the analytic function proposed in Ref.~\citenum{modelxe} for rare gas on metals.

The MD simulations performed in the second part of our
work show that at low temperatures and coverages the Xe atoms
deposited on the graphene cluster in small islands. The island
size increases with the temperature due to an increased particle
diffusion on the substrate. The island size increases also with
coverage until the Xe layer percolates and a unique patch of
coalesced islands is formed. We monitor the commensurability
of islands during their growth by both visual and quantitative
methods based on the calculated structure factor. The results
uncover the existence of a close correlation between the island
size and commensurability: small islands are in register with
the substrate, while larger islands are less commensurate.
The simulations repeated under equivalent conditions on gold
reveal a similar trend, but a much lower commensurability is
found for every considered temperature and coverage.

Considering the close relation between the static friction
and the interfacial commensurability~\cite{Reguzzoni2012}, our results can explain
the existence of a critical coverage for the depinning transition
and its dependence on temperature as observed in QCM
experiments~\cite{mistura}. Furthermore, they confirm the theory on the
size dependence of static friction, according to which nominal
incommensurate interfaces becomes commensurate below a
critical size~\cite{Reguzzoni2012}. The critical size, that depends on the interparticle
interaction strength and the potential corrugation, is dominated
by the lattice misfit. The latter dependence accounts for the
different frictional behavior observed for the Xe/GL and Xe/Au
systems.

\section{Acknowledgments}
We acknowledge the CINECA consortium for the availability of high performance computing
resources and support through the ISCRA-A ``Lubric'' project.

\bibliography{xeongraphene}%
\bibliographystyle{apsrev4-1}%

\end{document}